\documentclass[12pt]{article}
\usepackage{graphicx,epsfig}

\begin{document}

\title{\
Extremal properties of conditional entropy and quantum discord for XXZ, symmetric quantum states
}

\author{
M.~A.~Yurischev\footnote{E-mail:\ yur@itp.ac.ru}
}

\medskip
\date{\sl
Institute of Problems of Chemical Physics of the Russian Academy of Sciences,
142432 Chernogolovka, Moscow Region, Russia
}

\maketitle

\begin{abstract}
For the XXZ subclass of symmetric two-qubit X states,
we study the behavior of quantum conditional entropy $S_{cond}$ as a function of
measurement angle $\theta\in[0,\pi/2]$.
Numerical calculations show that the function $S_{cond}(\theta)$ for
X states can have at most one local extremum in the open interval from zero to $\pi/2$
(unimodality property).
If the extremum is a minimum the quantum discord displays region with variable
(state-dependent) optimal measurement angle $\theta^*$.
Such $\theta$-regions (phases, fractions) are very tiny in the space of X state
parameters.
We also discover the cases when the conditional entropy has a local {\em maximum}
inside the interval $(0,\pi/2)$.
It is remarkable that the maxima exist in surprisingly
wide regions and the boundaries for such regions are defined
by the same bifurcation conditions as for those with a minimum.
Moreover, the found maxima can exceed the conditional entropy values at the ends
of interval $[0,\pi/2]$ more than by $1\%$.
This instils hope in the possibility to detect such maxima in experiment.
\end{abstract}

\section{Introduction}
\label{sect:Intro}
Quantum correlations lie at the heart of quantum information science.
Different measures have been proposed to quantify them.
Owing to the crucial role in a number of quantum information problems and particularly
in the algorithm of deterministic quantum computation with one pure qubit (DQC1),
a special place among those measures belongs to the quantum discord
\cite{MBCPV12,Str15,ABC16,BT16}.

The notion of information-theoretic quantum discord, $Q$, for a bipartite system, $AB$,
has been introduced by Zurek in 2000 \cite{Z00}.
He defined the quantum discord as the difference between the quantum generalizations
of symmetric and asymmetric forms of classical mutual information.
This difference depends on the measurement which is performed
on a subsystem (say, $B$) to obtain the quantum conditional entropy $S_{cond}$
entering the asymmetric form of quantum mutual information.
As a result, the definition of quantum discord was reduced to the relation
$Q=S(\rho_B) - S(\rho_{AB}) + S_{cond}$,
where $S(\rho_{AB})$ and $S(\rho_B)$ are the entropies for the quantum states
$\rho_{AB}$ and $\rho_B$ of joint system $AB$ and its subsystem $B$, respectively.
To eliminate the undesirable dependence of quantum discord on the specific measurement,
one performs optimization over all measurements.
It may be a minimization or, v.v., maximization.
Taking into account physical (information-theoretic) reasoning,
the minimization over the complete set of local orthogonal-projective (von Neumann)
measurements $\{{\rm\Pi_i}\}$ has been taken
in the final definition of quantum discord \cite{OZ01,Z03}.
Another possible invariant quantity, namely the maximized discord, remained outside
the attention.

Notice, the orthogonal projectors $\{{\rm\Pi}_i=|i^\prime\rangle\langle i^\prime|\}$
can be parametrized by the two angles $\theta$ and $\phi$:
$|0^\prime\rangle=\cos{(\theta/2)}|0\rangle - e^{-i\phi}\sin{(\theta/2)}|1\rangle$ and
$|1^\prime\rangle=e^{i\phi}\sin{(\theta/2)}|0\rangle + \cos{(\theta/2)}|1\rangle$.
Thus, the optimization for the general two-qubit systems is reduced to that over
two variables.

Due to the optimization procedure, evaluation of quantum discord in practice entails
great difficulties even for the two-qubit systems.
Closed analytical formula up to the present has been derived only for the
Bell-diagonal states~\cite{Luo08}.

An attempt to extend the success of Luo \cite{Luo08} to the X states was undertaken
in 2010 by Ali, Rau, and Alber \cite{ARA10}.
Unfortunately, the authors decided that the extreme values of parameters which
characterize the von Neumann measurement are attained only at their end points.
Shortly after, however, the counterexamples of X density matrices have been
given which demonstrate a conditional entropy minimum inside
the interval of measurement parameters \cite{LMXW11,CZYYO11}.
Thus, the analytic formula of Ref.~\cite{ARA10} is incorrect in general.

At that time it was also established that for X states the optimization
of conditional entropy (and hence discord) over the projectors
$\{{\rm\Pi}_i(\theta,\phi)\}$ can be worked out exactly over the azimuthal angle
$\phi$ but one optimization procedure, in the polar angle $\theta\in[0,\pi/2]$,
remains relevant \cite{LXSW10,CRC10,VR12}.
As a result, a pessimistic verdict has been made:
``For general two-qubit X states quantum discord cannot be evaluated
analytically'' \cite{H13} (see also \cite{MHR15}\footnote{
Unfortunately, the authors of Ref.~\cite{MHR15} have made a few mistakes.
In particular, their equation~(23) is incorrect and their conclusion about the
existence of the region with mysterious measurement $\sigma_?$ is wrong.}
and \cite{JY16}).

Definite optimism was restored in Refs.~\cite{Y14,Y14a,Y15}, where
it has been observed that the formula for calculating the quantum discord
of two-qubit X states has, in any event, a piecewise-analytical-numerical
(semianalytical) form
\begin{equation}
   \label{eq:Q3}
   Q=\min\{Q_0, Q_{\theta^*}, Q_{\pi/2}\}.
\end{equation}
Here the subfunctions (branches) $Q_0$ and $Q_{\pi/2}$ are the analytical expressions
(corresponding to the discord with optimal measurement angles 0 and $\pi/2$, respectively)
and only the third branch $Q_{\theta^*}$ requires one-dimensional searching of the
optimal state-dependent measurement angle $\theta^*\in(0,\pi/2)$ if, of course,
the interior global extremum exists.
Notice, a similar situation with the variable optimal angle
can also occur for the one way-quantum deficit of two-qubit X states \cite{YWF16}.

Three possible regions (phases, fractions) $Q_0$, $Q_{\pi/2}$, and $Q_{\theta^*}$ are
separated in the density-matrix parameter space by sharp demarcation boundaries.
An equation for the boundary between $Q_0$ and $Q_{\pi/2}$ regions is,
obviously, $Q_0=Q_{\pi/2}$ or $S_{cond}(0)=S_{cond}(\pi/2)$.
The equations for 0- and $\pi/2$-boundaries separating respectively $Q_0$ and $Q_{\pi/2}$
regions with the $Q_{\theta^*}$ one have the forms \cite{Y14,Y14a,Y15}
\begin{equation}
   \label{eq:SII1}
   S^{\prime\prime}_{cond}(0)=0, \qquad   S^{\prime\prime}_{cond}(\pi/2)=0,
\end{equation}
where $S^{\prime\prime}_{cond}(0)$ and $S^{\prime\prime}_{cond}(\pi/2)$ are the
second derivatives of function $S_{cond}(\theta)$ with respect to  $\theta$
at the end points $\theta=0$ and $\pi/2$.

Equations~(\ref{eq:SII1}) are based on the idea
of bifurcation mechanism \cite{A04} for appearance and disappearance
of intermediate local extremum: by smooth varying the parameters of density matrix
elements the interior extremum of $S_{cond}(\theta)$ can come into the interval
$(0,\pi/2)$ or come out of it only through the end points $\theta=0,\pi/2$.
Bifurcation mechanism is supported by the following two properties of
function $S_{cond}(\theta)$.
On the one hand, its first derivatives at $\theta=0$ and $\pi/2$
identically equal zero, $S^\prime_{cond}(0)\equiv S^\prime_{cond}(\pi/2)\equiv0$.
This follows from direct calculations for the function $S_{cond}(\theta)$
which is known in general X-state case (\cite{Y15} and references therein).
On the other hand, we suppose the unimodal property for the function $S_{cond}(\theta)$
(see Appendix).
So, if the unimodality hypothesis is valid the only possibility
(except the trivial case $S_{cond}(\theta)=const$) for a single local extremum
(minimum or maximum) to appear or disappear inside the open interval by continuous
varying the parameters defining the X state is the doubling the extremun
at the ends of interval $[0,\pi/2]$.
Such a bifurcation mechanism for sudden birth (creation) and sudden death (annihilation)
of interior extremum  has been proposed by the author in Refs.~\cite{Y14,Y14a,Y15}.

Once both equations (\ref{eq:SII1}) have been solved and it was found
that the boundaries do not coincide,
one should check the shape of curve $S_{cond}(\theta)$ inside the corridor found.
If the curve exhibits the interior local extremum hence this will prove that the region
under question exists in fact.
Moreover, by this we can distinguish if the region contains the minimum or maximum.

The remainder of this article is arranged as follows.
Domain of discord function is given in Sec.~\ref{sect:Domain},
complete three-dimensional discord phase diagram is presented in Sec.~\ref{sect:Phase},
regions with the intermediate minimum and maximum of conditional entropy
are found, respectively, in Secs.~\ref{sect:Min} and \ref{sect:Max}, and
some proposals for possible experiments are discussed in Sec.~\ref{sect:Propos}.
The results obtained are briefly summarized in Sec.~\ref{sect:Concl}.
Finally, the Appendix includes material concerning the unimodal functions.

\section{Density matrix and domain of definition for its entries}
\label{sect:Domain}
Density matrix $\rho_{AB}$ of general two-qubit X state contains seven independent
parameters.
However, it is well known (see, e.g, \cite{Y14}) that such a density matrix
can be reduced by using local unitary transformations to the real five-parameter form
\begin{equation}
   \label{eq:rhoXr-Bloch}
   \rho_{AB}=\frac{1}{4}(1
   + s_1\sigma_z\otimes1
   + s_21\otimes\sigma_z
   + c_1\sigma_x\otimes\sigma_x 
   + c_2\sigma_y\otimes\sigma_y 
   + c_3\sigma_z\otimes\sigma_z),
\end{equation}
where $\sigma_\alpha$ ($\alpha=x,y,z$) are the Pauli matrices.
The coefficients $s_1, s_2, c_1, c_2, c_3$ are the unary and binary correlation
functions, i.e., experimentally measurable quantities.
It is clear that
\begin{equation}
   \label{eq:s1c3}
   -1\le s_1, s_2, c_1, c_2, c_3\le 1.
\end{equation}

The domain of definition, ${\cal D}$, in the five-dimensional space spanned by
$(s_1, s_2, c_1, c_2, c_3)$ is formed according to the requirement of
density-matrix positive semidefiniteness which leads to the conditions
\cite{CZYYO11,Y15,KHJP10} (see also \cite{NJ13})
\begin{equation}
   \label{eq:surfaces}
   (1 - c_3)^2\ge(s_1 - s_2)^2 + (c_1 + c_2)^2,\quad
   (1 + c_3)^2\ge(s_1 + s_2)^2 + (c_1 - c_2)^2. 
\end{equation}
The solid ${\cal D}$ is finite,
lies in the five-dimensional hypercube $[-1,1]^5$ in conforming with Eq.~(\ref{eq:s1c3}),
results from an intersection of two conic hypercylinders \cite{Y15} in accord
with Eq.~(\ref{eq:surfaces}), and its volume is about 8\% of the hypercube one.

Having taken $10,000$ randomly chosen two-qubit X states the authors~\cite{VR12}
found that $\approx99.8\%$ of such states belong to the phases $Q_0$ and $Q_{\pi/2}$.
On the other hand, specific examples show that there are $Q_{\theta^*}$-regions with sizes
of order $10^{-4}$ and less 
(see below Sec.~\ref{sect:Min} and Fig.~\ref{fig:z034dtr}).
Therefore, to inspect the five-dimensional domain ${\cal D}$ with such a resolution,
the running of a fivefold nested loop on hypothetical 10~GHz processor\footnote{
At present,
the maximum clock frequency of
Intel and AMD microprocessors is around
8.7-8.8~GHz.}
will require the time more than a thousand years!

To avoid the arisen difficulty one can build up an atlas (collection) of ''maps``.
In other words, it is reasonable to study different particular families of total
X-state family.
Here, we take an important in practice subclass of X states and perform for it
a careful investigation.
But before, let us consider one classical example to extract useful inferences
from it.

\subsection{Luo's formula as an instructive example}
\label{subsect:Luo}
\begin{figure}[t]
\begin{center}
\epsfig{file=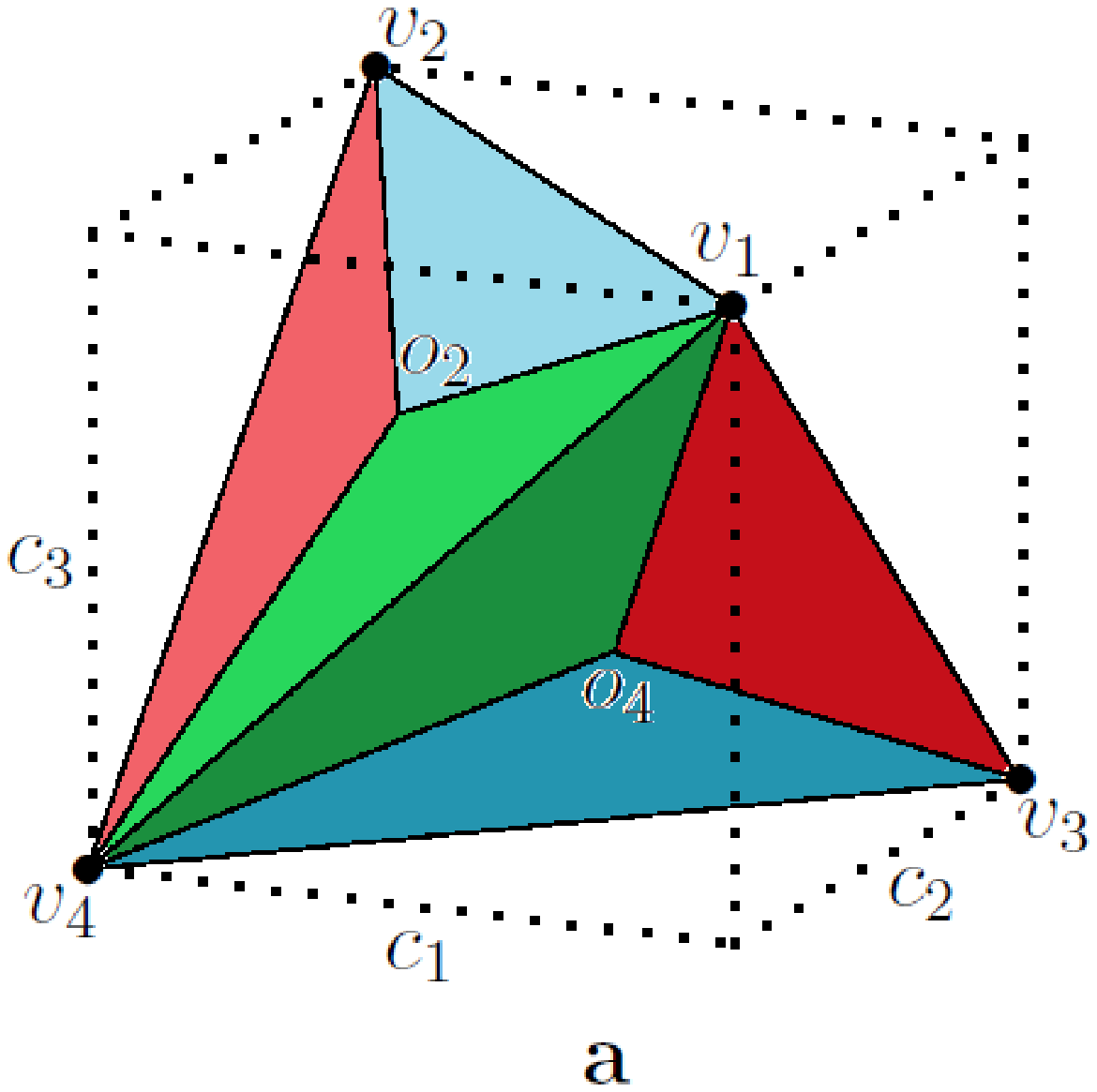,width=5.6cm}
\hspace{0.3cm}
\epsfig{file=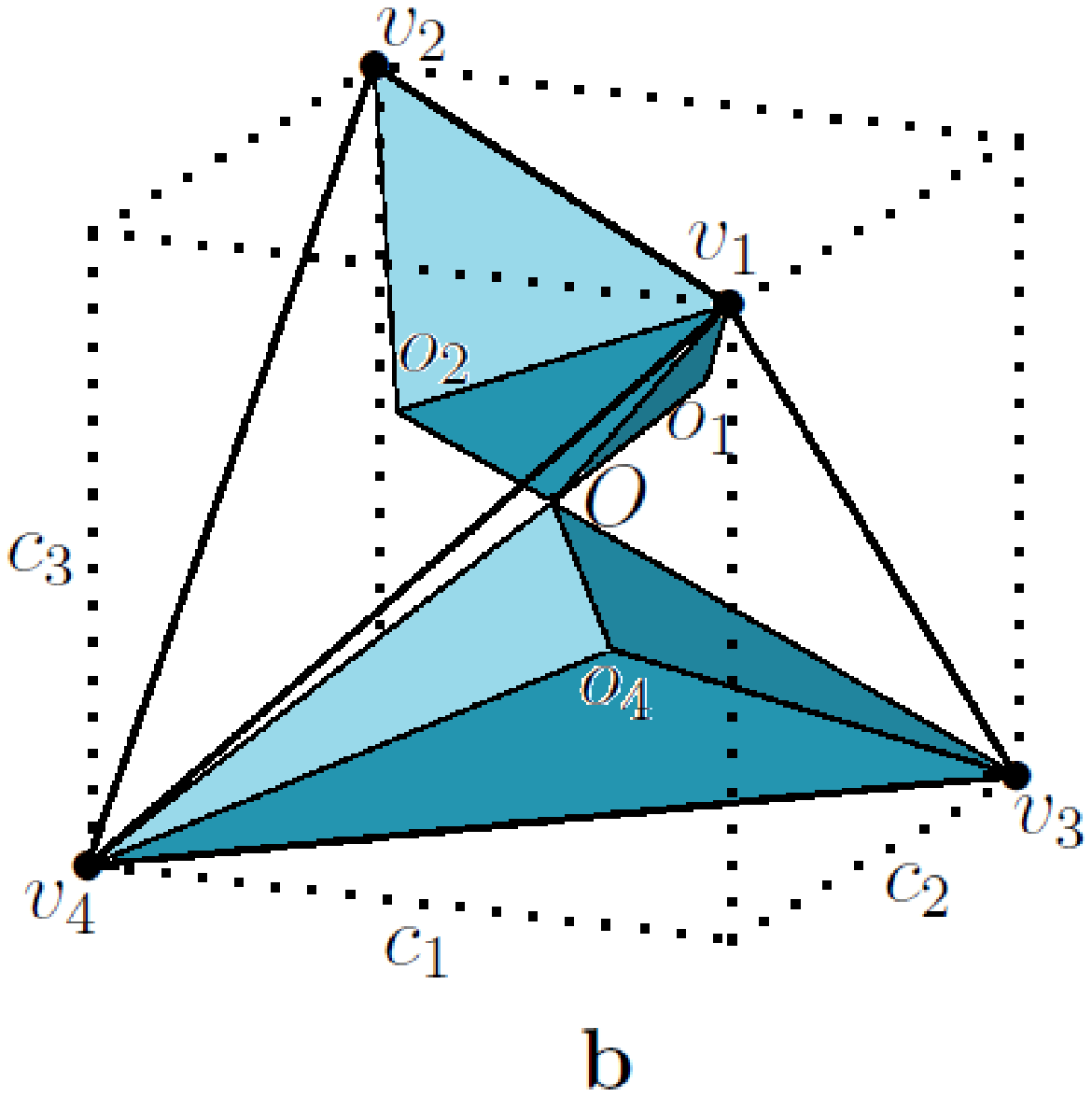,width=5.6cm}
\caption{(Color online)
Phase diagram for the discord function of Bell-diagonal states.
The red, green, and blue regions are the domains of definition respectively
for the three branches $Q_1$, $Q_2$, and $Q_3$ of discord function.
Here, (${\bf a}$) is an outward appearance of tetrahedron $v_1v_2v_3v_4$ and
(${\bf b}$) shows the $Q_3$ region;
regions $Q_1$ and $Q_2$ are similar but include the pairs
of edges $\{v_1v_3,v_2v_4\}$ and $\{v_1v_4,v_2v_3\}$, correspondingly
}
\label{fig:tetrBd}
\end{center}
\end{figure}

A particular case $s_1=s_2=0$ (i.e., when both local Bloch vectors are zero)
corresponds to the Bell-diagonal states
\begin{equation}
   \label{eq:rho-Bell}
   \rho_{AB}=\frac{1}{4}(I
   + c_1\sigma_x\otimes\sigma_x 
   + c_2\sigma_y\otimes\sigma_y 
   + c_3\sigma_z\otimes\sigma_z).
\end{equation}
Eigenvalues of this density matrix equal
$p_1=(1+c_1-c_2+c_3)/4$, $p_2=(1-c_1+c_2+c_3)/4$, 
$p_3=(1+c_1+c_2-c_3)/4$, and $p_4=(1-c_1-c_2-c_3)/4$. 
Due to the nonnegativity definition of density operators, all $p_i\ge0$
and therefore physical values of parameters $c_1$, $c_2$, and $c_3$
lie in the perfect tetrahedron which is inscribed in the three-dimensional
cube $[-1,1]^3$ and has the vertices
$v_1=(1,-1,1)$, $v_2=(-1,1,1)$, $v_3=(1,1,-1)$, and $v_4=(-1,-1,-1)$ \cite{HH96};
see Fig.~\ref{fig:tetrBd}.
Faces of this tetrahedron are equilateral triangles and their
centers are $o_1=(1/3,1/3,1/3)$, $o_2=(-1/3,-1/3,1/3)$,
$o_3=(-1/3,1/3,-1/3)$, and $o_4=(1/3,-1/3,-1/3)$.
Volume of this body equals a third of the $[-1,1]^3$ cube volume.

The quantum discord of Bell-diagonal states is given by Luo's formula \cite{Luo08},
which can be written as
\begin{equation}
   \label{eq:LuoQ}
   Q=\min\{Q_1,Q_2,Q_3\}, 
\end{equation}
where
\begin{equation}
   \label{eq:Qi}
   Q_i=\sum_{k=1}^4p_k\log_2(4p_k)-\frac{1}{2}[(1-c_i)\log_2(1-c_i)+(1+c_i)\log_2(1+c_i)]
\end{equation}
for $i=1,2,3$.
Luo's formula is perfect and has the same strong proof as, for example,
Pythagoras' theorem.

Three possible cases $|c_1|\ge\max\{|c_2|,|c_3|\}$, $|c_2|\ge\max\{|c_1|,|c_3|\}$, 
and $|c_3|\ge\max\{|c_1|,|c_2|\}$ lead, as shown in Fig.~\ref{fig:tetrBd},
to a separation of tetrahedron by diagonal planes
$|c_1|=|c_2|$, $|c_2|=|c_3|$, and $|c_3|=|c_1|$ respectively into three domains
$Q_1$, $Q_2$, and $Q_3$ \cite{Y15,MGY17}.
Discord reaches maximal values (equal one in bit units) at the tetrahedron
vertices and vanishes on the Cartesian axes $Oc_1$, $Oc_2$, and $Oc_3$.

Pay attention that the quantum discord (\ref{eq:LuoQ}) is a
piecewise-defined function.
Because of this, when the system passes from one subdomain to another,
the discord at the boundary experiences a sudden (abrupt) change
which can be displayed as a fracture on its curve
(see, e.g., a recent review \cite{CM16}).
This is similar to the first-order phase transitions which occur in liquids
or solids.

In the case of Bell-diagonal states, the 0- and $\pi/2$-boundaries
coincide because both equations (\ref{eq:SII1})
are reduced to one relation $2|c_3|=|c_1+c_2|+|c_1-c_2|$, that is equivalent to
four equations $c_3=\pm c_1$ and $c_3=\pm c_2$ (see Ref.~\cite{Y15}).
These planes divide the tetrahedron into subdomains $Q_0$ and $Q_{\pi/2}$.
The $Q_0$-subdomain consists of two hexahedrons $(O,v_1,v_2,o_1,o_2)$ and
$(O,v_3,v_4,o_3,o_4)$;
they are shown in Fig.~\ref{fig:tetrBd}(b).
Remaining volume of the tetrahedron belongs to the $Q_{\pi/2}$ states.
It is twice as large as the volume of $Q_0$ region.

Because the 0- and $\pi/2$-boundaries are coincident,
$Q_{\theta^*}$-subdomain is absent and quantum discord is given by an
explicit analytical formula $Q=\min\{Q_0,Q_{\pi/2}\}$ which is in full
agreement with Eqs.~(\ref{eq:LuoQ}) and (\ref{eq:Qi}).
This becomes obvious if we take into account that the branch $Q_{\pi/2}(c_1,c_2,c_3)$
contains the piecewise functions like $|x|$ and therefore splits, in turn,
into two subbranches, $Q_{\pi/2}^{(x)}$ and $Q_{\pi/2}^{(y)}$.
As a result, the formula takes the form (see \cite{Y15})
\begin{equation}
   \label{eq:QBd1}
   Q=\min\{Q_{\pi/2}^{(x)},Q_{\pi/2}^{(y)},Q_0\},
\end{equation}
where the branches $Q_{\pi/2}^{(x)}$, $Q_{\pi/2}^{(y)}$, and $Q_0$
correspond, respectively, to $Q_1$, $Q_2$, and $Q_3$ phases.

So, we learn useful lessons from the Bell-diagonal case and Luo's formula:
the discord function is a piecewise one
(this results, obviously, from the optimization condition)
and the piecewise structure may be nested (a hierarchy of piecewise functions).

\subsection{Domain for the XXZ-model with parity symmetry}
\label{subsect:XXZS}
Let us consider a two-qubit system with the three-parameter X density matrix
\begin{equation}
   \label{eq:rhoXXZS}
   \rho_{AB}
	 =\frac{1}{4}\!\left(
      \begin{array}{cccc}
      1+2s_1+c_3&0&0&0\\
      0&1-c_3&2c_1&0\\
      0&2c_1&1-c_3&0\\
      0&0&0&1-2s_1+c_3
      \end{array}
   \right)
\end{equation}
or in the Bloch form
\begin{equation}
   \label{eq:rhoXXZS-Bloch}
   \rho_{AB}=\frac{1}{4}[1
   + s_1(\sigma_z\otimes1
   + 1\otimes\sigma_z)
   + c_1(\sigma_x\otimes\sigma_x 
   + \sigma_y\otimes\sigma_y) 
   + c_3\sigma_z\otimes\sigma_z].
\end{equation}
Thus, the general X state (\ref{eq:rhoXr-Bloch}) is restricted here by conditions
$c_2=c_1$ and $s_2=s_1$.
Such a model encounters often in different physically important problems.

\begin{figure}[t]
\begin{center}
\epsfig{file=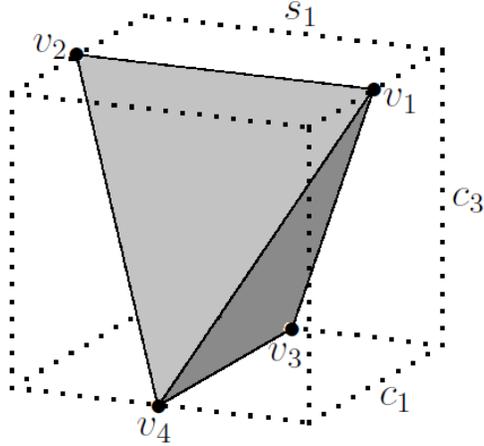,width=8cm}
\caption{
Shaded tetrahedron ${\cal T}$ embedded in three-dimensional cube
(dotted lines) is the domain of discord function $Q(s_1,c_1,c_3)$ ---
``Bloch sphere'' for states (\ref{eq:rhoXXZS}) or (\ref{eq:rhoXXZS-Bloch})
}
\label{fig:z_xxz-m2}
\end{center}
\end{figure}

Eigenvalues of matrix (\ref{eq:rhoXXZS}) are equal to
$\lambda_{1,2}=(1\pm2s_1+c_3)/4$ and $\lambda_{3,4}=(1\pm2c_1-c_3)/4$.
From the requirement $\rho_{AB}\succeq0$, it follows that all $\lambda_j\ge0$
($j=1,\ldots,4$).
As a result, the domain of definition of the discord function $Q(s_1,c_1,c_3)$
is a tetrahedron ${\cal T}$ which is defined as
\begin{equation}
   \label{eq:c3s1c1}
   c_3\in[-1,1],\quad
   s_1\in[-(1+c_3)/2,(1+c_3)/2],\quad c_1\in[-(1-c_3)/2,(1-c_3)/2].
\end{equation}
This tetrahedron is enclosed in the three-dimensional cube $[-1,1]^3$,
has the vertices $v_1$, $v_2$, $v_3$, and $v_4$ and
isosceles triangle faces (see Fig.~\ref{fig:z_xxz-m2}).
Volume of tetrahedron ${\cal T}$ equals one sixth part of cube volume ($=2^3$).
So, one may say that the discord $Q(s_1,c_1,c_3)$ is a function on the tetrahedron ${\cal T}$.

It is easy to check that the matrices of structure (\ref{eq:rhoXXZS})
with different values of parameters,
$\{s_1,c_1,c_3\}$ and $\{s_1^\prime,c_1^\prime,c_3^\prime\}$, commute.
Hence the fidelity, $F$, between two states $\rho_{AB}$ and  $\rho_{AB}^\prime$
can be calculated by simple formula
\begin{equation}
   \label{eq:fid}
   F=\biggl(\sum_{j=1}^4\sqrt{\lambda_j\lambda_j^\prime}\biggr)^2,
\end{equation}
where $\lambda_j^\prime$ are the eigenvalues of matrix $\rho_{AB}^\prime$
with a form (\ref{eq:rhoXXZS})
but with parameters $s_1^\prime$, $c_1^\prime$, and $c_3^\prime$.
Fidelity is known to characterize a measure of closeness and distinguishability
of quantum states.

The density matrix under discussion, Eqs.~(\ref{eq:rhoXXZS}) or (\ref{eq:rhoXXZS-Bloch}),
can be map into the Gibbs density matrix $\rho$ of the thermalized XXZ spin dimer in
a homogeneous magnetic field applied in $z$-direction,
\begin{equation}
   \label{eq:rhoG}
   \rho=\frac{1}{Z}\exp(-H/T).
\end{equation}
Here $T$ is the temperature in energy units, $Z$ is the partition function,
and the Hamiltonian reads
\begin{equation}
   \label{eq:H}
   H=-\frac{1}{2}[J(\sigma_x^1\sigma_x^2+\sigma_y^1\sigma_y^2)+J_z\sigma_z^1\sigma_z^2]
	   -\frac{1}{2}B(\sigma_z^1+\sigma_z^2),
\end{equation}
where $J$ and $J_z$ are the exchange coupling constants and $B$ is the external field.
The energy levels of dimer equal
\begin{equation}
   \label{eq:Ei}
   E_{1,2}=-\frac{1}{2}(J_z\pm 2B),\qquad E_{3,4}=-\frac{1}{2}(J_z\pm 2J)
\end{equation}
and therefore the partition function is given by
\begin{equation}
   \label{eq:Z}
   Z=2\biggl(e^{J_z/2T}\cosh\frac{B}{T}+e^{-J_z/2T}\cosh\frac{J}{T}\biggl).
\end{equation}
Parameters of density matrix (\ref{eq:rhoXXZS-Bloch}) equal
\begin{eqnarray}
   \label{eq:s1c1c3JJzTB}
   &&s_1=\frac{2}{Z}e^{J_z/2T}\sinh\frac{B}{T},\quad
	 c_1=\frac{2}{Z}e^{-J_z/2T}\sinh\frac{J}{T},
	 \nonumber\\
   &&\quad c_3=\frac{2}{Z}\biggl(e^{J_z/2T}\cosh\frac{B}{T}-e^{-J_z/2T}\cosh\frac{J}{T}\biggl).
\end{eqnarray}
Inversely, three parameters $J$, $J_z$, and $B$ are expressed through
the density matrix parameters $s_1$, $c_1$, $c_3$ and the temperature $T$
as follows
\begin{equation}
   \label{eq:JJzB}
   J=\frac{T}{2}\ln\frac{1+2c_1-c_3}{1-2c_1-c_3},\quad
   J_z=\frac{T}{2}\ln\frac{(1+c_3)^2-4s_1^2}{(1-c_3)^2-4c_1^2},\quad
   B=\frac{T}{2}\ln\frac{1+2s_1+c_3}{1-2s_1+c_3}.
\end{equation}

Thus, to gain an insight into the physical reasons
we can translate different effects into the physical language of spin Hamiltonian
as a function of temperature and magnetic field by different values
of anisotropy parameter $\Delta=J_z/|J|$.
The above relations allow to map $\{s_1,c_1,c_3\}\to\{T/|J|,B/|J|,\Delta\}$.
Such a spin-model language is useful, e.g., in investigation of a spin detection
efficiency of nonideal ferromagnetic detectors \cite{KGBTSM14,RBKTGBSM15,MG16}.

\section{Branches $Q_0$ and $Q_{\pi/2}$. Phase diagram}
\label{sect:Phase}
The whole domain ${\cal T}$ of discord function $Q(s_1,c_1,c_3)$ has been found
in the previous section.
Our next task is to determine the subdomains of ${\cal T}$ which answer different
fractions of discord.
We will solve this problem in two stages.
As a preliminary, the tetrahedron ${\cal T}$ will be divided into regions corresponding
to the phases of pseudo-discord
\begin{equation}
   \label{eq:Q-tilde}
    {\tilde Q}=\min\{Q_0,Q_{\pi/2}\}.
\end{equation}
Thereafter, in the next section, the phase $Q_{\theta^*}$ will be located in
the solid ${\cal T}$.

Expressions for $Q_0$ and $Q_{\pi/2}$ as functions of matrix elements
of the general X density matrix
can be found in Refs.~\cite{H13,MHR15,Y14,Y14a,Y15}\footnote{
In Eq.~(19) of Ref.~\cite{Y15} the common factor of $\frac{1}{4}$
in the expression for $\Lambda_{1,2}$
should be replaced by $\frac{1}{2}$ (as in Refs.~\cite{Y14,Y14a})}.
Our case of interest is $s_2=s_1$ and $c_2=c_1$.
In this particular 
case, quantum discord with the optimal measurement angle
$\theta=0$ is given by
\begin{eqnarray}
   \label{eq:Q0s1c1c3}
   &&Q_0(s_1,c_1,c_3)=\frac{1}{4}[-2(1-c_3)\ln(1-c_3)
	 \nonumber\\
	 &&+(1+2c_1-c_3)\ln(1+2c_1-c_3) + (1-2c_1-c_3)\ln(1-2c_1-c_3)].
\end{eqnarray}
(Here, the discord is in nat measurement units;
to transform the discord in bit units, one should divide it by $\ln2$.)
The function (\ref{eq:Q0s1c1c3}) does not depend on $s_1$, it is even
with respect to $c_1$ and identically equals zero by $c_1=0$
(i.e., in the classical limit).

For the branch $Q_{\pi/2}$, we have
\begin{eqnarray}
   \label{eq:Qpi/2s1c1c3}
   &&Q_{\pi/2}(s_1,c_1,c_3)=-\frac{1}{2}\biggl[(1+s_1)\ln(1+s_1) + (1-s_1)\ln(1-s_1)
	 \nonumber\\
	 &&+\biggl(1+\sqrt{s_1^2+c_1^2}\biggr)\ln\biggl(1+\sqrt{s_1^2+c_1^2}\biggr)
	 +\biggl(1-\sqrt{s_1^2+c_1^2}\biggr)\ln\biggl(1-\sqrt{s_1^2+c_1^2}\biggr)\biggr]
	 \nonumber\\
	 &&+\frac{1}{4}[(1+2c_1-c_3)\ln(1+2c_1-c_3) + (1-2c_1-c_3)\ln(1-2c_1-c_3)
	 \nonumber\\
	 &&+ (1+2s_1+c_3)\ln(1+2s_1+c_3) + (1-2s_1+c_3)\ln(1-2s_1+c_3)].
\end{eqnarray}
This function is symmetric both on $s_1$ and $c_1$.

Let us break up the tetrahedron ${\cal T}$ into the regions $Q_0$ and $Q_{\pi/2}$
temporarily ignoring a possible existence of region $Q_{\theta^*}$.
In this case, the boundaries between the regions $Q_0$ and $Q_{\pi/2}$
are defined by the condition  $Q_0=Q_{\pi/2}$ and, taken into account Eqs.~(\ref{eq:Q0s1c1c3})
and ({\ref{eq:Qpi/2s1c1c3}), can be found
from the transcendental equation
\begin{eqnarray}
   \label{eq:Qpi2s1c1c3}
	 &&\qquad\quad 2\bigl[-(1-c_3)\ln(1-c_3) + (1+s_1)\ln(1+s_1) + (1-s_1)\ln(1-s_1)
	 \nonumber\\
	 &&+ \biggl(1+\sqrt{s_1^2+c_1^2}\biggr)\ln\biggl(1+\sqrt{s_1^2+c_1^2}\biggr)
	 + \biggl(1-\sqrt{s_1^2+c_1^2}\biggr)\ln\biggl(1-\sqrt{s_1^2+c_1^2}\biggr)\bigr]
	 \nonumber\\
	 &&\quad -(1+2s_1+c_3)\ln(1+2s_1+c_3) - (1-2s_1+c_3)\ln(1-2s_1+c_3)=0.
\end{eqnarray}
If $c_3$ is fixed then the boundaries will be symmetric under
mirror reflections $s_1\to -s_1$ and $c_1\to -c_1$.

We will now slice up the tetrahedron ${\cal T}$ by planes with different
fixed values of $c_3\in[-1,1]$.
By this, two other parameters, $s_1$ and $c_1$, may vary according to
Eq.~(\ref{eq:c3s1c1}) in the limits which depend only on $c_3$.
Let $c_3=-1$, then $s_1=0$ and therefore
we are moving on the edge $v_3v_4$ (see Fig.~\ref{fig:z_xxz-m2}).
In this case, the equation (\ref{eq:Qpi2s1c1c3}) leads to the solutions $c_1=\pm1$.
On the edge $v_3v_4$, the values of $Q_0$ are less than those of $Q_{\pi/2}=\ln2$ nat
(i.e., one bit) and hence
\begin{equation}
   \label{eq:Q0v3v4}
   Q=\frac{1}{2}[(1+c_1)\ln(1+c_1) + (1-c_1)\ln(1-c_1)].
\end{equation}
It will be shown in the next section that the fraction $Q_{\theta^*}$ is
absent when $c_3\le0$ and therefore this expression yields true values for the discord.
Thus, the discord is one bit at the vertices $v_3$ and $v_4$ and varies
according to Eq.~(\ref{eq:Q0v3v4}) between them
reaching zero at the middle of edge $v_3v_4$, i.e., when $c_1=0$.

If $c_3$ somewhat increases then the cross section of tetrahedron ${\cal T}$
takes the shape of a stretched rectangle at the two ends of which the regions $Q_{\pi/2}$
are located and the remaining flat of the rectangle belongs to the region $Q_0$.
Phase diagram of discord at $c_3=-0.5$ is shown in Fig.~\ref{fig:ph-d}(a).
\begin{figure}[t]
\begin{center}
\epsfig{file=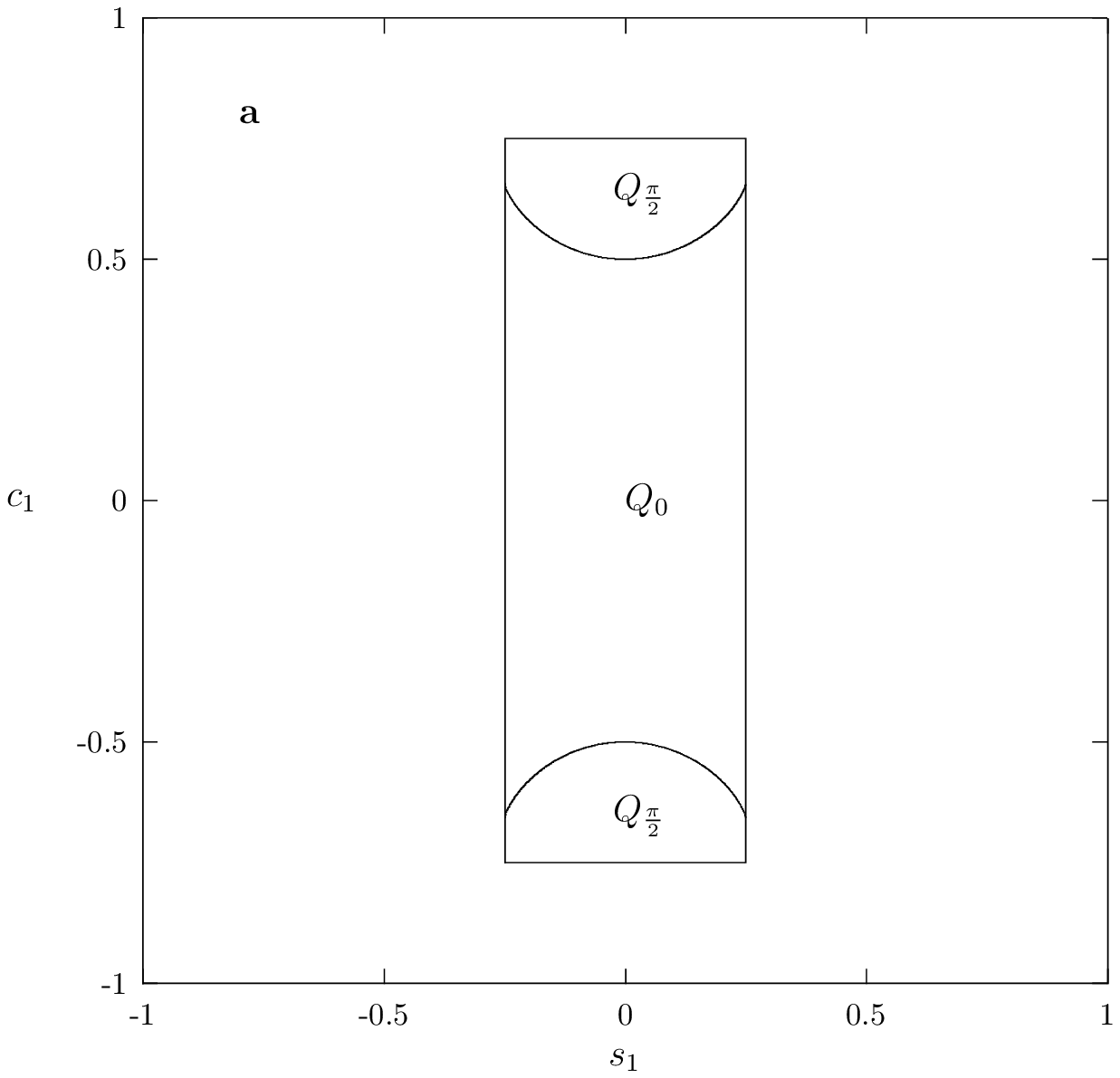,width=5.2cm}
\vspace{1cm}
\hspace{1cm}
\epsfig{file=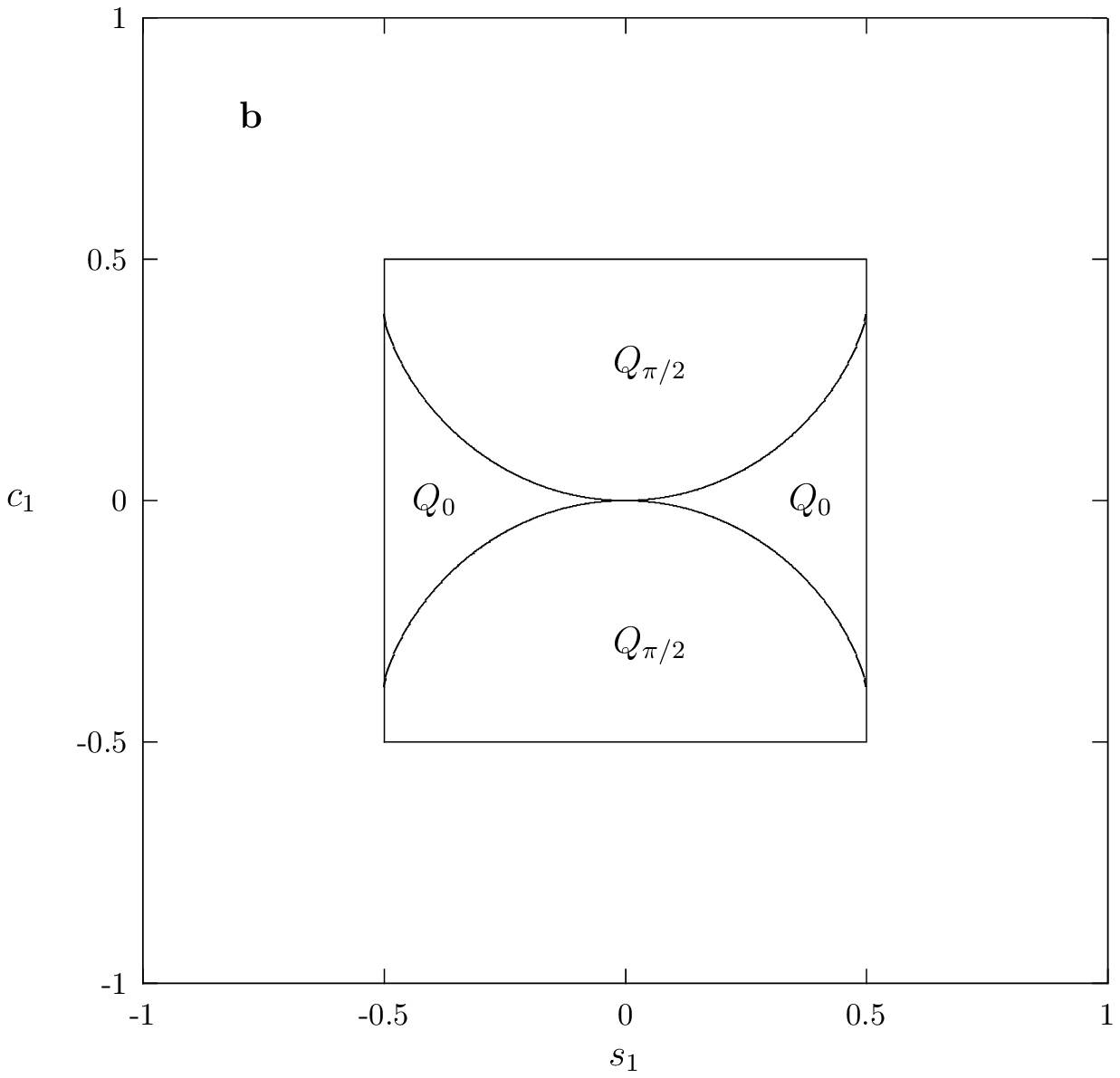,width=5.2cm}
\epsfig{file=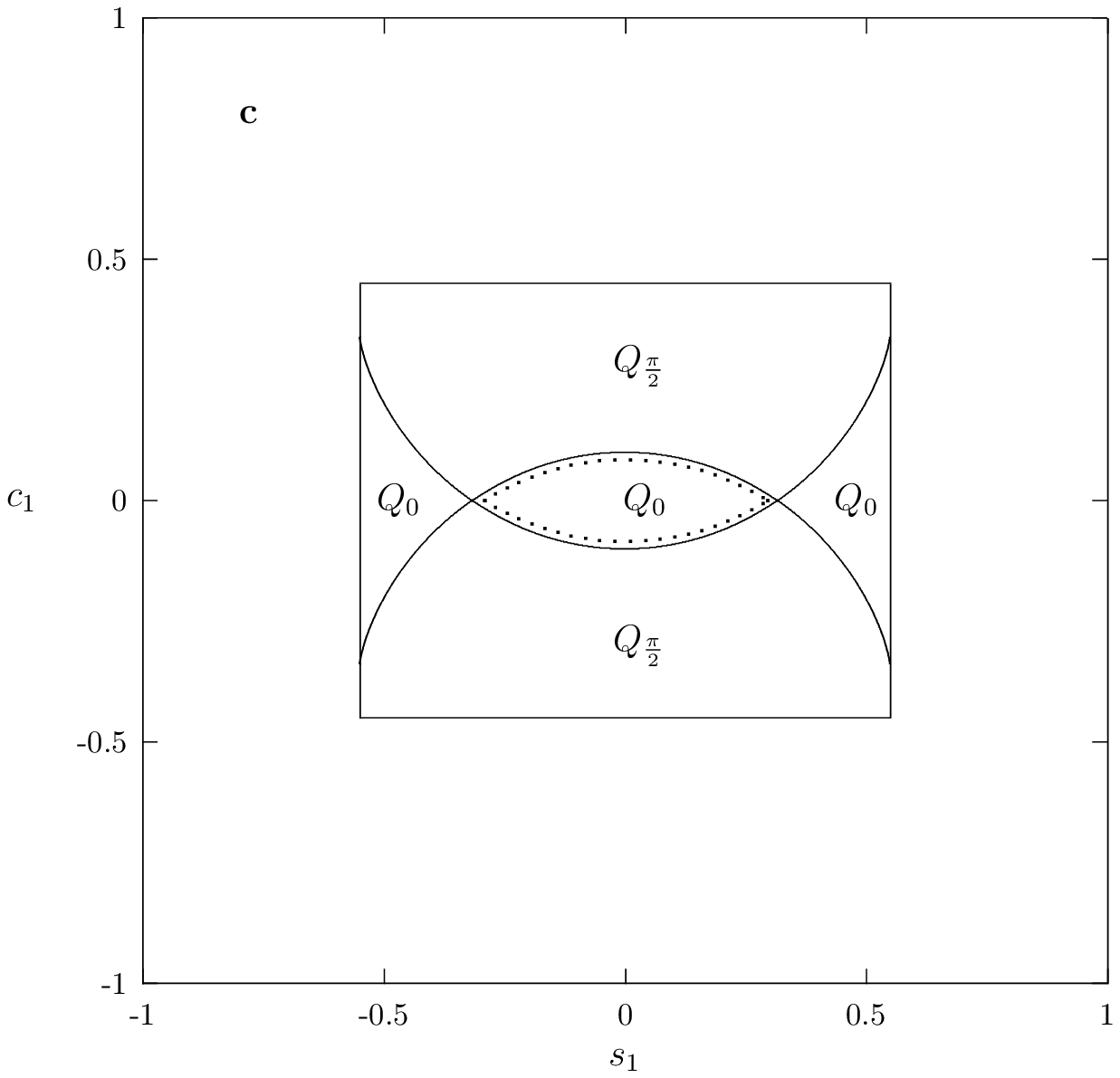,width=5.2cm}
\vspace{1cm}
\hspace{1cm}
\epsfig{file=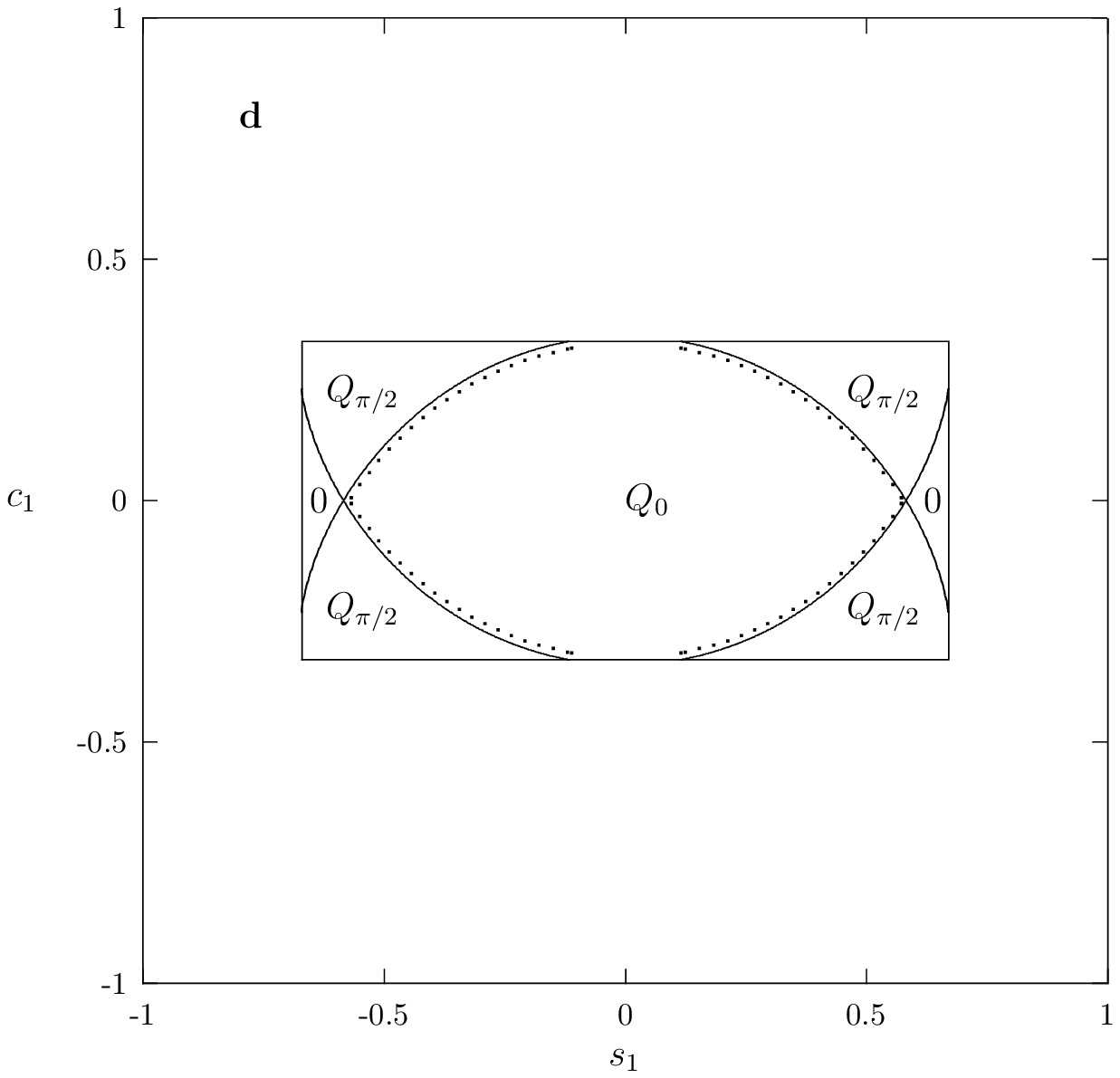,width=5.2cm}
\caption{
Phase diagrams for the discord by different sections of tetrahedron ${\cal T}$
by planes $c_3=-0.5$~(a), $0$~(b), $0.1$~(c), and  $0.34$~(d).
(On the section (d), the region 0 is the same as $Q_0$.)
Boundaries between the phases $Q_0$ and $Q_{\pi/2}$ are shown by single solid lines.
The $Q_{\theta^*}$ regions are thin too and cannot be resolved in the given scale;
their positions are marked by double solid-dotted lines, see (c) and (d) 
}
\label{fig:ph-d}
\end{center}
\end{figure}
Two boundaries between the regions $Q_{\pi/2}$ and $Q_0$ are some arcs
which cross the $Oc_1$ axis (i.e., when $s_1=0$) at the points $c_1=\pm c_3$. 
Notice for a reference that in the case under discussion ($c_3=-0.5$),
the values of $c_1$ on the arcs at $s_1=\pm 0.25$ equal $\pm 0.656\,390\,9127$
(see Fig.~\ref{fig:ph-d}(a)).

By further increasing the height of tetrahedron section, two subregions
$Q_{\pi/2}$ will move towards each other and meet by $c_3=0$
at the point $s_1=c_1=0$.
This situation is fixed in Fig.~\ref{fig:ph-d}(b).
Here the cross section of tetrahedron ${\cal T}$ is a square.

When $c_3$ reaches positive values, the cross section will be
a rectangle stretched now in the $s_1$-direction.
Remarkably, the new part appeared between the boundaries of two subregions
$Q_{\pi/2}$ is occupied by the phase $Q_0$.
This is depicted in Fig.~\ref{fig:ph-d}(c) for $c_3=0.1$.
Here, the arcs are intersected on the axis $Os_1$ at the points
$s_1=\pm0.316\,227\,7647$.

With further increasing of the values of $c_3$, the new appeared (inner) region
$Q_0$ will increase and by $c_3=1/3$ the arcs touch the horizontal sides of the
rectangle.
Thereafter the region $Q_{\pi/2}$ will consist of four $Q_{\pi/2}$-subregions
isolated in the cross section.
This situation is shown in Fig.~\ref{fig:ph-d}(d) by $c_3=0.34$.
Here, the arc-like curves are intersected at $s_1=\pm0.583\,095\,1892$,
reach the vertical sides ($s_1=\pm0.67$) of the rectangle at
$c_1=\pm0.231\,128\,4073$, and the horizontal ones ($c_1=\pm0.33$) at
$s_1=\pm0.113\,209\,2068$.

In the limit $c_3\to1$, the $Q_0$-phase becomes again the only dominant on
the edge $v_1v_2$ but here the discord tends to zero because $c_1$ vanishes.

From the above analysis it is not difficult to reconstruct
a three-dimensional picture --- the solid $Q_{\pi/2}$ (or, v.v., $Q_0$)
inside the tetrahedron ${\cal T}$.
An outward appearance of tetrahedron ${\cal T}$ and corresponding regions
are shown in Fig.~\ref{fig:z_xxzb2a}.
\begin{figure}[t]
\begin{center}
\epsfig{file=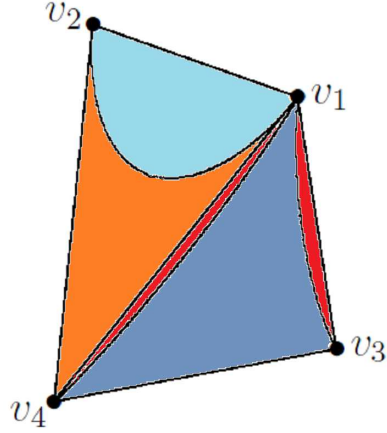,width=6cm}
\caption{(Color online)
An outward appearance of tetrahedron ${\cal T}$ with the regions $Q_0$ (blue)
and $Q_{\pi/2}$ (orange/red).
The region $Q_{\theta^*}$ lies inside the tetrahedron and therefore is not seen
}
\label{fig:z_xxzb2a}
\end{center}
\end{figure}

In the next sections, it will be shown that the phase diagrams presented
in Fig.~\ref{fig:ph-d} are true for $c_3\le0$
while when $c_3>0$ the inner, newly appeared region $Q_0$ and the $Q_{\pi/2}$
ones are separated by very thin layers which are not resolved in the scale
of Fig.~\ref{fig:ph-d} and therefore their locations are shown schematically
by the double solid-dotted lines.

\section{Regions with the interior {\em minimum} of conditional entropy}
\label{sect:Min}
To find the phase $Q_{\theta^*}$ we will study the behavior of conditional
entropy shapes.
Using again the general expression for the average quantum conditional entropy
of five-parameter X state \cite{Y14,Y14a,Y15}, we obtain that in case of interest
$s_2=s_1$ and $c_2=c_1$ the entropy is given as
\begin{eqnarray}
   \label{eq:Sconds1c1c3}
   &&S_{cond}(\theta; s_1,c_1,c_3)
	 \nonumber\\
	 &&=\ln2+\frac{1}{2}[(1+s_1\cos\theta)\ln(1+s_1\cos\theta) + (1-s_1\cos\theta)\ln(1-s_1\cos\theta)]
	 \nonumber\\
	 &&-\frac{1}{4}\biggl[\biggl(1+s_1\cos\theta+\sqrt{(s_1+c_3\cos\theta)^2+c_1^2\sin^2\theta}\biggr)\ln\biggl(1+s_1\cos\theta
	 \nonumber\\
	 &&\qquad +\sqrt{(s_1+c_3\cos\theta)^2+c_1^2\sin^2\theta}\biggr)
	 \nonumber\\
	 &&+\biggl(1+s_1\cos\theta-\sqrt{(s_1+c_3\cos\theta)^2+c_1^2\sin^2\theta}\biggr)\ln\biggl(1+s_1\cos\theta
	 \nonumber\\
	 &&\qquad -\sqrt{(s_1+c_3\cos\theta)^2+c_1^2\sin^2\theta}\biggr)
	 \nonumber\\
	 &&+\biggl(1-s_1\cos\theta+\sqrt{(s_1-c_3\cos\theta)^2+c_1^2\sin^2\theta}\biggr)\ln\biggl(1-s_1\cos\theta
	 \nonumber\\
	 &&\qquad -\sqrt{(s_1-c_3\cos\theta)^2+c_1^2\sin^2\theta}\biggr)
	 \nonumber\\
	 &&+\biggl(1-s_1\cos\theta-\sqrt{(s_1-c_3\cos\theta)^2+c_1^2\sin^2\theta}\biggr)\ln\biggl(1-s_1\cos\theta
	 \nonumber\\
	 &&\qquad -\sqrt{(s_1-c_3\cos\theta)^2+c_1^2\sin^2\theta}\biggr)\biggr].
\end{eqnarray}
This function is differentiable at any point $\theta$ and, 
in full conferment with the statement made in the Introduction,
its  first derivatives with respect to $\theta$ identically
equal zero for $\forall\,s_1,c_1,c_3\in{\cal T}$
at both ends of the interval $[0,\pi/2]$,
\begin{equation}
   \label{eq:d1Scond0}
   S^\prime_{cond}(0;s_1,c_1,c_3)\equiv0,\qquad S^\prime_{cond}(\pi/2;s_1,c_1,c_3)\equiv0.
\end{equation}

Let us investigate different types of conditional entropy behavior
by various values of $s_1$, $c_1$, and $c_3$ belonging to the tetrahedron ${\cal T}$.
For this purpose, we consider, for definiteness, a motion in the plane $c_3=0.34$
along the strait line $c_1=0.14$ when $s_1$ varies from zero to its maximal value
$(1+c_3)/2$~($=0.67$).
This trajectory is shown by dotted horizontal line in Fig.~\ref{fig:z034trs}.
\begin{figure}[t]
\begin{center}
\epsfig{file=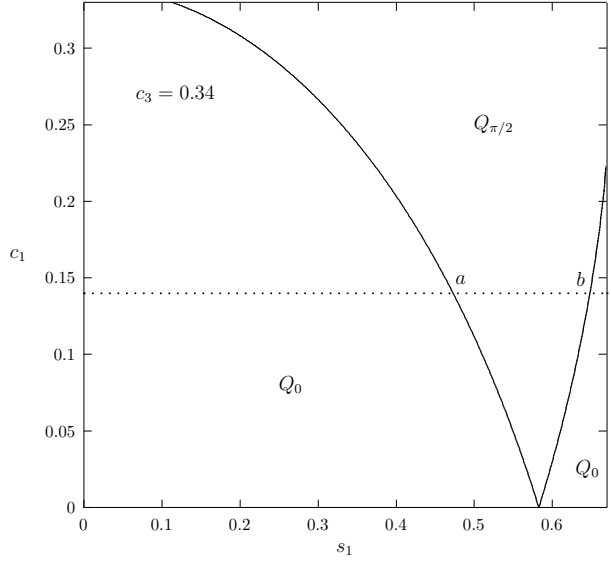,width=8cm}
\caption{
Quarter of a phase diagram by $c_3=0.34$.
The dotted straight line $c_1=0.14$ crosses the 
($Q_0$-$Q_{\pi/2}$)-boundaries
at the points $a$ and $b$ in the vicinities of which the conditional entropy
$S_{cond}(\theta)$ displays the intermediate minimum and maximum, respectively
}
\label{fig:z034trs}
\end{center}
\end{figure}
It intersects the boundaries defined by equation $Q_0=Q_{\pi/2}$
at two points: $s_1^a=0.473\,267$ and $s_1^b=0.648\,435$ (see Fig.~\ref{fig:z034trs}).
When the values of $s_1$ are small enough, the curve of $S_{cond}(\theta)$ has monotonically
increasing shape, i.e., its minimum lies at $\theta=0$ and hence the quantum
discord equals $Q_0$.
However, in the neighborhood of point $s_1^a$ the curve is deformed and
the intermediate minimum is arisen as depicted in Fig.~\ref{fig:s-a}(a)-(b).
\begin{figure}[t]
\begin{center}
\epsfig{file=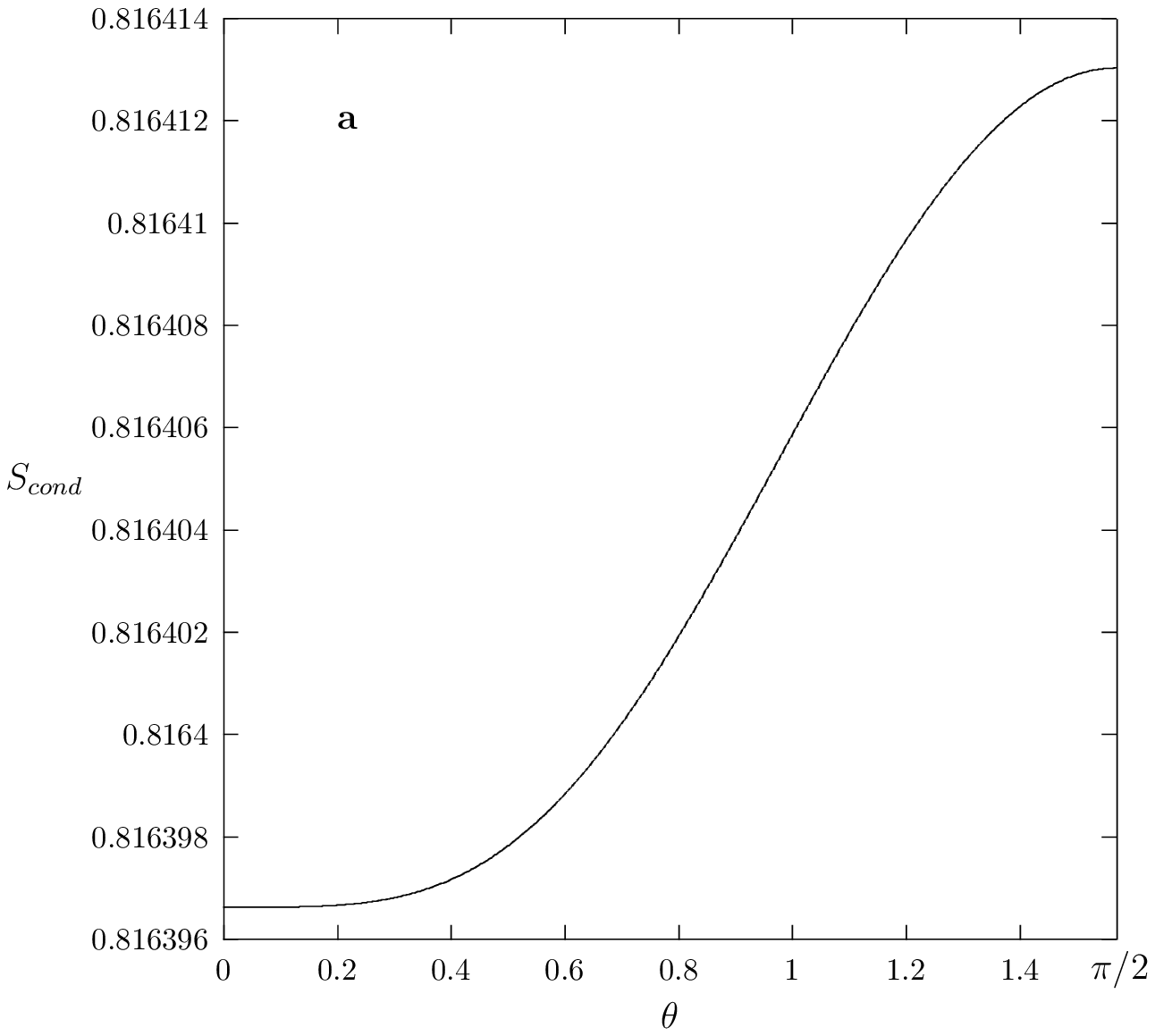,width=5.2cm}
\vspace{1cm}
\hspace{1cm}
\epsfig{file=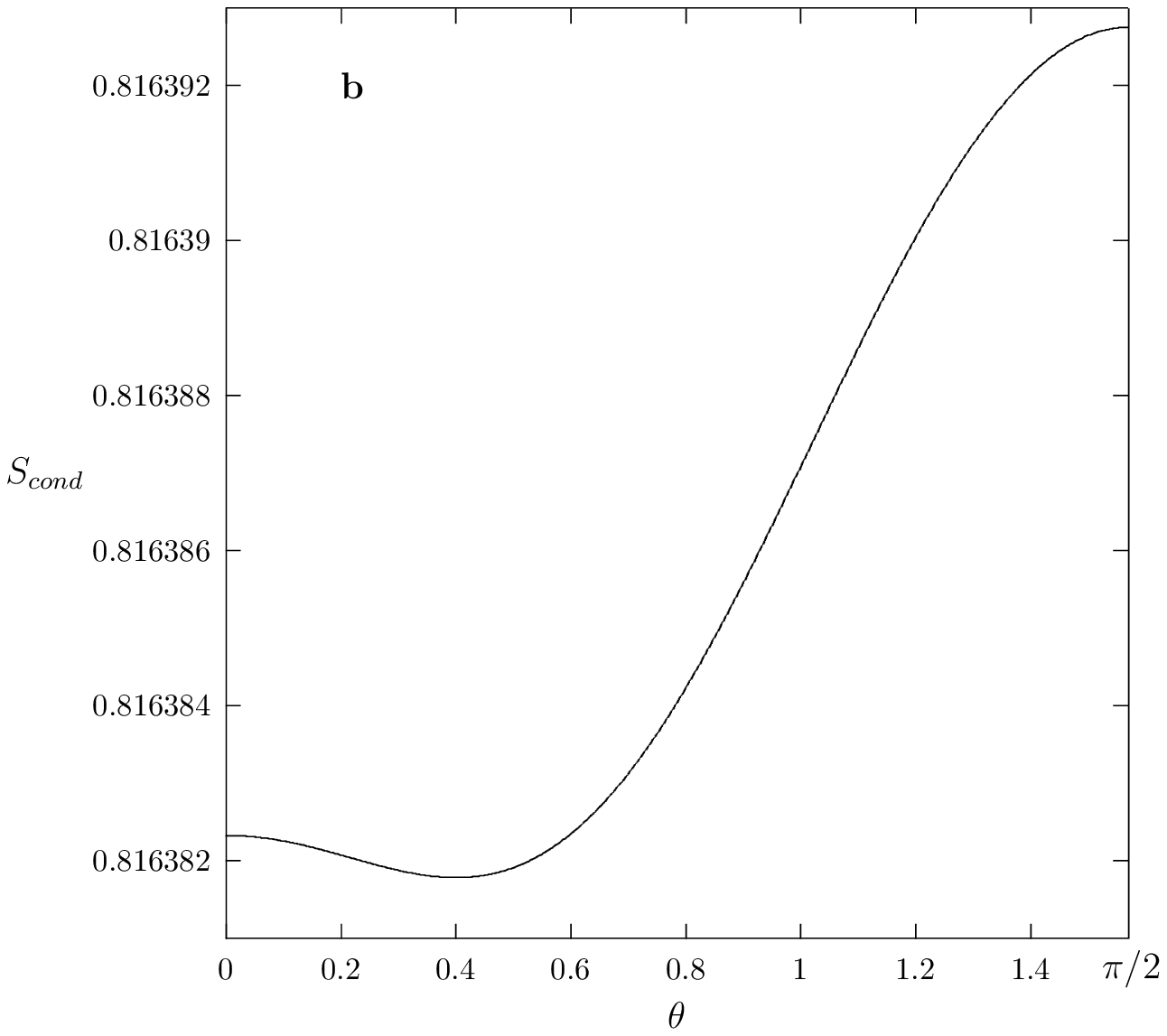,width=5.2cm}
\epsfig{file=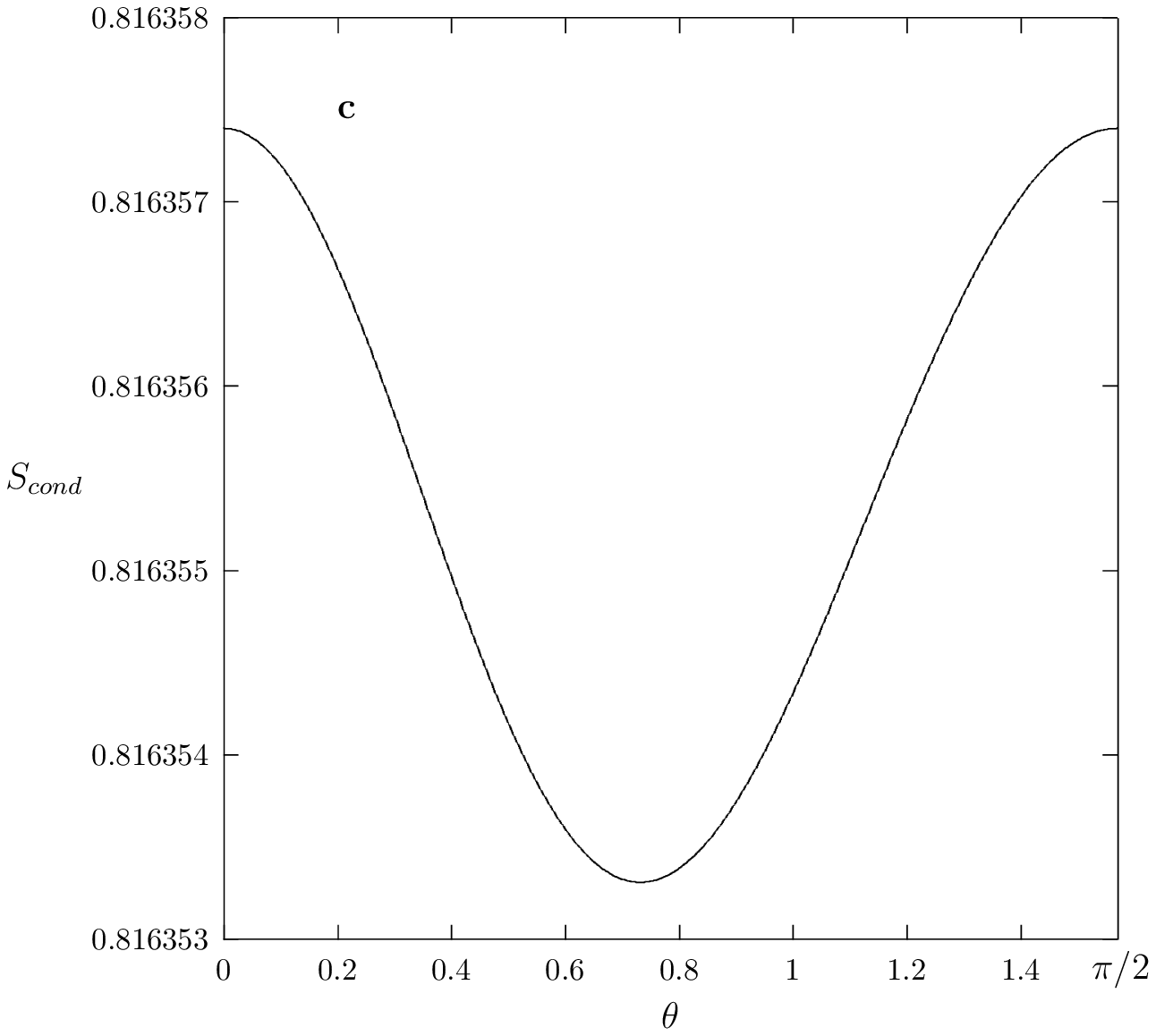,width=5.2cm}
\vspace{1cm}
\hspace{1cm}
\epsfig{file=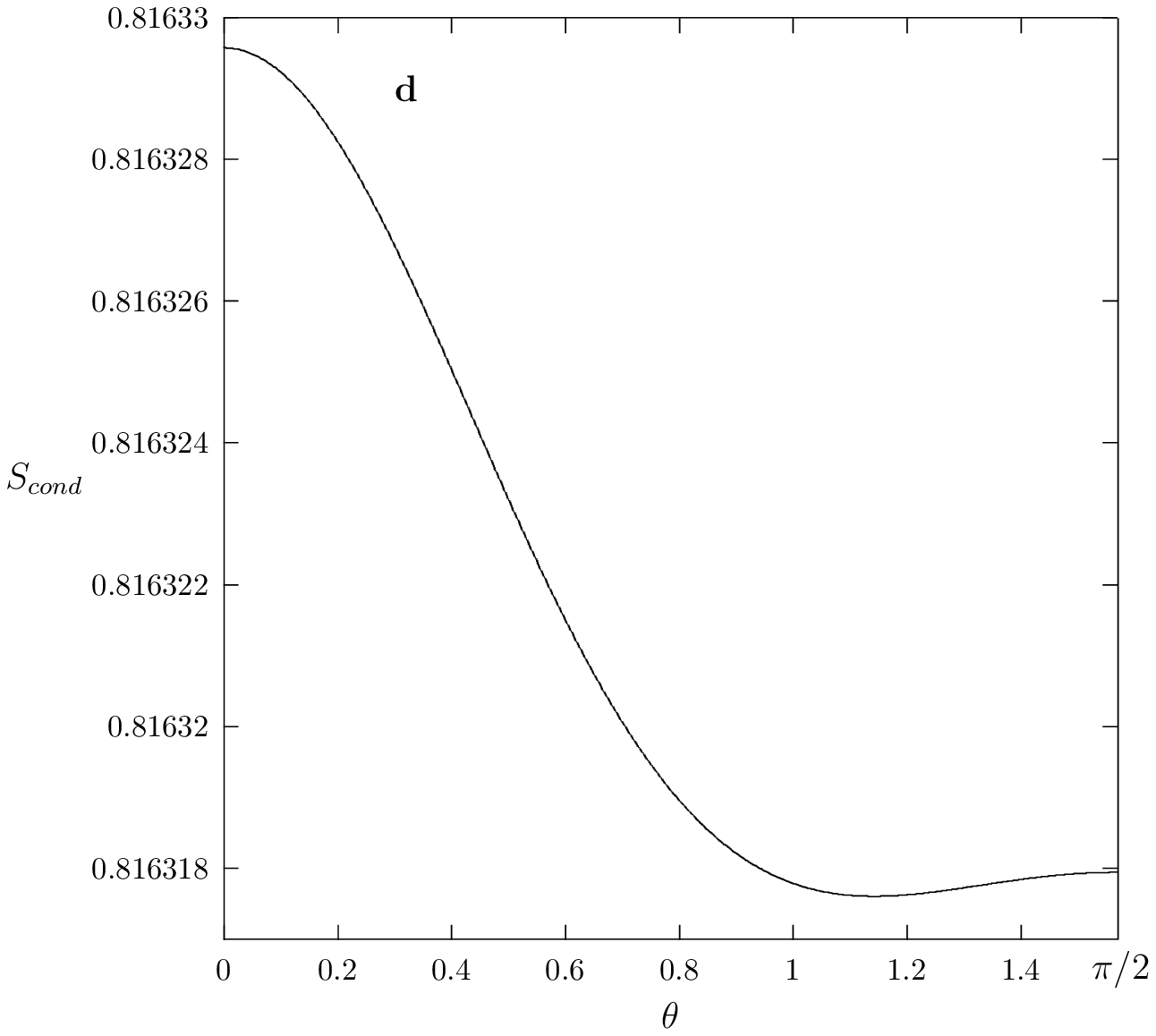,width=5.2cm}
\caption{
Conditional entropy $S_{cond}$ (in bit units) vs measurement angle $\theta$
by $c_3=0.34$, $c_1=0.14$, and $s_1=0.473\,193$~(a), 0.473\,22~(b), 0.473\,267~(c),
0.473\,32~(d).
Here, the sudden birth, life, and sudden death of $S_{cond}$-minimum are
observed
}
\label{fig:s-a}
\end{center}
\end{figure}
The largest depth of minimum is achieved near by $s_1=s_1^a$
(see Fig.~\ref{fig:s-a}(c)).
Here, $S_{cond}(0)=S_{cond}(\pi/2)=0.816\,357\,3993$~bit and
$S_{cond}^{min}=0.816\,353\,3082$~bit
which is achieved at $\theta=0.732\,419\approx42^o$.
Thus, the decrease is $\Delta S_{cond}=-0.000\,004\,09$~bit, i.e.,
relative decreasing equals $0.0005\%$ only.

Thanks to the property (\ref{eq:d1Scond0}) and because the minimum is single
(which confirms the unimodality hypothesis), the minimum could come
into the open interval $(0,\pi/2)$ by continuous variation of state parameters
only from the end point.
Such a mechanism of bearing the interior minimum through a bifurcation (its doubling)
is illustrated in Fig.~\ref{fig:zs034014-b1} in wider window $-\pi\le\theta\le\pi$.
\begin{figure}[t]
\begin{center}
\epsfig{file=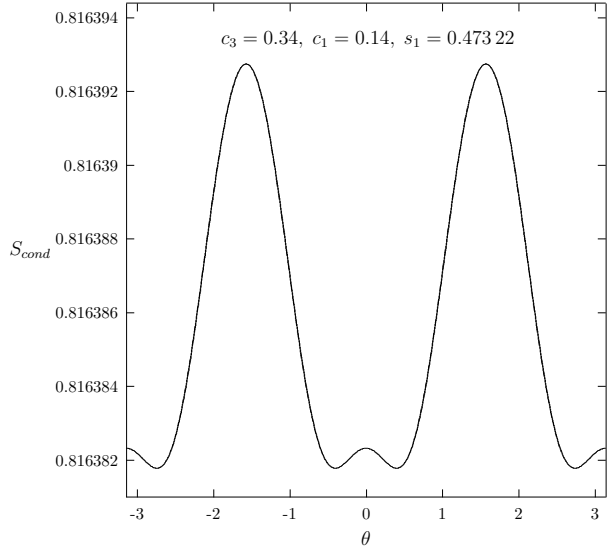,width=8cm}
\caption{
Conditional entropy $S_{cond}$ (in bits) vs $\theta$ in the band $[-\pi,\pi]$
by $c_3=0.34$, $c_1=0.14$, and $s_1=0.473\,22$.
This graph demonstrates the bifurcation of minimum at $\theta=0$
}
\label{fig:zs034014-b1}
\end{center}
\end{figure}

So, it is naturally to accept the unimodality hypothesis which
allows to determine the sharp boundaries from Eqs.~(\ref{eq:SII1}).
Using Eq.~(\ref{eq:Sconds1c1c3}) or general expressions from Refs.~\cite{Y14,Y14a,Y15} we
arrived at
\begin{eqnarray}
   \label{eq:S2prime0}
	 &&S_{cond}^{\prime\prime}(0;s_1,c_1,c_3)
	 \nonumber\\
	 &&=\frac{1}{4}\biggl[-c_1^2\bigg(\frac{1}{s_1+c_3}\ln\frac{1+2s_1+c_3}{1-c_3}
	 + \frac{1}{s_1-c_3}    \ln\frac{1-c_3}{1-2s_1+c_3}\biggl)
	 \nonumber\\
	 &&s_1\biggl(2\ln\frac{1-s_1}{1+s_1}+\ln\frac{1+2s_1+c_3}{1-2s_1+c_3}\biggr)
	 +c_3\ln\frac{(1+2s_1+c_3)(1-2s_1+c_3)}{(1-c_3)^2}\biggr]\quad
\end{eqnarray}
and
\begin{eqnarray}
   \label{eq:S2primepi2}
	 &&S_{cond}^{\prime\prime}(\pi/2;s_1,c_1,c_3)
	 =s_1^2 + \frac{c_1^2(s_1^2+c_1^2-c_3^2)}{2(s_1^2+c_1^2)^{3/2}}\ln\frac{1+\sqrt{s_1^2+c_1^2}}{1-\sqrt{s_1^2+c_1^2}}
	 \nonumber\\
	 &&-\frac{s_1^2}{2}\biggl[\frac{1}{1+\sqrt{s_1^2+c_1^2}}\bigg(1+\frac{c_3}{\sqrt{s_1^2+c_1^2}}\biggr)^2
	 +\frac{1}{1-\sqrt{s_1^2+c_1^2}}\bigg(1-\frac{c_3}{\sqrt{s_1^2+c_1^2}}\biggr)^2\biggr].\qquad
\end{eqnarray}
Solving
Eqs.~(\ref{eq:SII1}) with these expressions 
we obtain the boundaries on the trajectory for the region $Q_{\theta^*}$:
it exists when $s_1\in[0.473\,192\,8814, 0.473\,341\,2570]$.
The width of given interval, $1.484\times10^{-4}$, is 
very small.
Fidelity between the boundary states of found $s_1$-interval
is $F=99.999\,998\%$, i.e., extremely high and therefore at present this
$Q_{\theta^*}$ region cannot be detected experimentally.

In Fig.~\ref{fig:z034dtr},
the deviations $\Delta_0=c_1^0-c_1^{\times}$ and $\Delta_{\pi/2}=c_1^{\pi/2}-c_1^{\times}$
as functions of $s_1$ are drawn, where $c_1^{\times}$ is the crossing point of the
branches $Q_0$ and $Q_{\pi/2}$ (i.e., a solution of equation $Q_0=Q_{\pi/2}$)
and $c_1^0$ and $c_1^{\pi/2}$ are the 0- and $\pi/2$-boundaries, i.e., the solutions
of equations $S_{cond}^{\prime\prime}(0)=0$ and $S_{cond}^{\prime\prime}(\pi/2)=0$,
respectively.
\begin{figure}[t]
\begin{center}
\epsfig{file=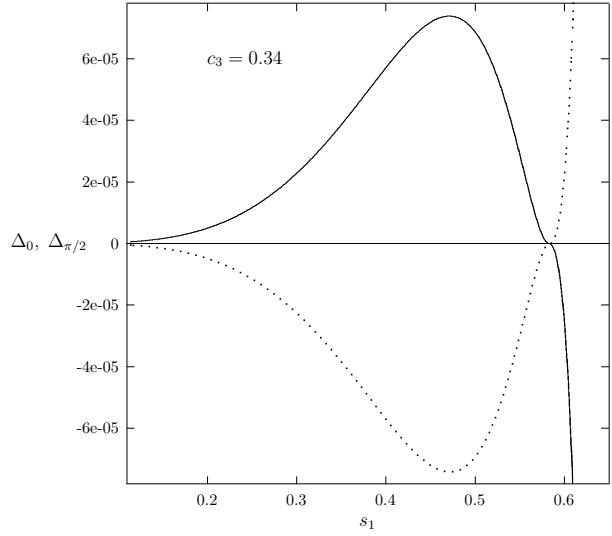,width=8cm}
\caption{
Deviations
$\Delta_0$ (dotted line) and
$\Delta_{\pi/2}$ (solid line) vs $s_1$ in the cross section $c_3=0.34$
}
\label{fig:z034dtr}
\end{center}
\end{figure}
Deviations vanish at $s_1=0.113\,209\,2068$ and $s_1=0.583\,095\,1892$;
their maximum magnitudes lie at $s_1=0.471\,198$
and equal $\Delta_0^{max}=-0.000\,074\,1487$ and
$\Delta_{\pi/2}^{max}=0.000\,073\,7956$.
Thus, the maximum width of $Q_{\theta^*}$ region in the $c_1$-direction is
$1.479\times10^{-4}$ that is very tiny again.

It is interesting to understand what the appearance and disappearance of
intermediate region $Q_{\theta^*}$ in the language of spin dimer at the thermal
equilibrium means; Eqs.~(\ref{eq:H})--(\ref{eq:JJzB}).
Taking $c_1=0.14$ and $c_3=0.34$ one gets $T/J=2.111\,082\,372$ for any $s_1$.
When $s_1$ varies in the $Q_{\theta^*}$ region, that is from $s_1^0=0.473\,192\,8814$ to
$s_1^{\pi/2}=0.473\,341\,2570$,
then the band for $\Delta=J_z/J$ lies between $1.019\,558\,9945$ and $1.020\,248\,4171$,
i.e., the width of interval is now equal to $6.894\,23\times10^{-4}$.
At the same time, normalized magnetic field $B/J$ varies from $1.942\,519\,04$ to
$1.953\,495\,0495$.
These results show that the phase $Q_{\theta^*}$ exists in a rather ordinary
ferromagnetic region of dimer (both constants $J$ and $J_z$ are positive) and
its sizes are very small as before.

On the other hand, taking $\Delta=1.02$ (from the above found interval)
and, for example, $B/J=1$ the calculations yield
that the conditional entropy has the intermediate minimum for $T/J$ in the interval
$[0.761\,06,0.853\,61]$ which relative width equals $11.5\%$ \cite{Y14,Y15}.
So, the same $Q_{\theta^*}$ region can has large sizes on the phase diagram
in $(T,B)$-variables by fixed $\Delta$.
This surprising result we interpret as follows.
The phase $Q_{\theta^*}$ is a layer separating the $Q_0$- and $Q_{\pi/2}$-phases
in the density matrix space $(s_1,c_1,c_3)$.
This layer is thin in one direction only.
Hence, if one considers the intermediate phase $Q_{\theta^*}$ in other directions
the sizes may be large.
However, even smallest fluctuations of system parameters (temperature,
magnetic field, etc.) can lead away the phase $Q_{\theta^*}$.

\section{Regions with the interior {\em maximum} of conditional entropy}
\label{sect:Max}
We continue to go on the plane $c_3=0.34$ in the straightforward way $c_1=0.14$.
In the interval where the phase $Q_{\theta^*}$ exists, the
optimal measurement angle smoothly changes from zero to $\pi/2$ as can be seen
from Fig.~\ref{fig:s-a}.
Therefore, when $s_1>0.473\,341\,2570$, the discord equals $Q_{\pi/2}$ down to
the neighborhood of point $b$ (see Fig.~\ref{fig:z034trs}).
Near the point $b$ a new phenomenon is observed.
Namely, the extremum appears again but now instead of minimum a single maximum is
arisen inside the interval $(0,\pi/2)$.
This situation is shown in Fig.~\ref{fig:z034tr_b}.
\begin{figure}[t]
\begin{center}
\epsfig{file=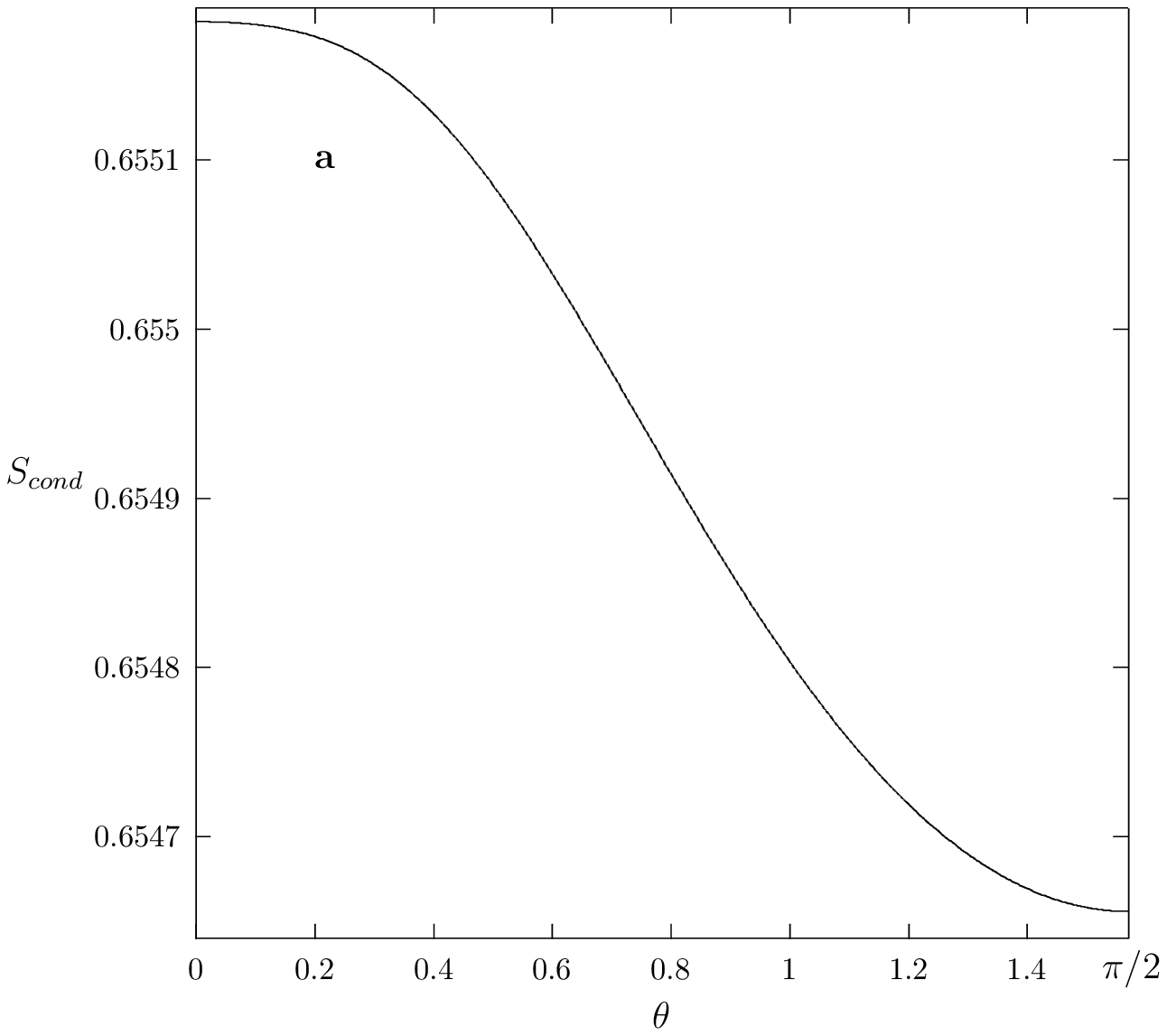,width=5.2cm}
\vspace{1cm}
\hspace{1cm}
\epsfig{file=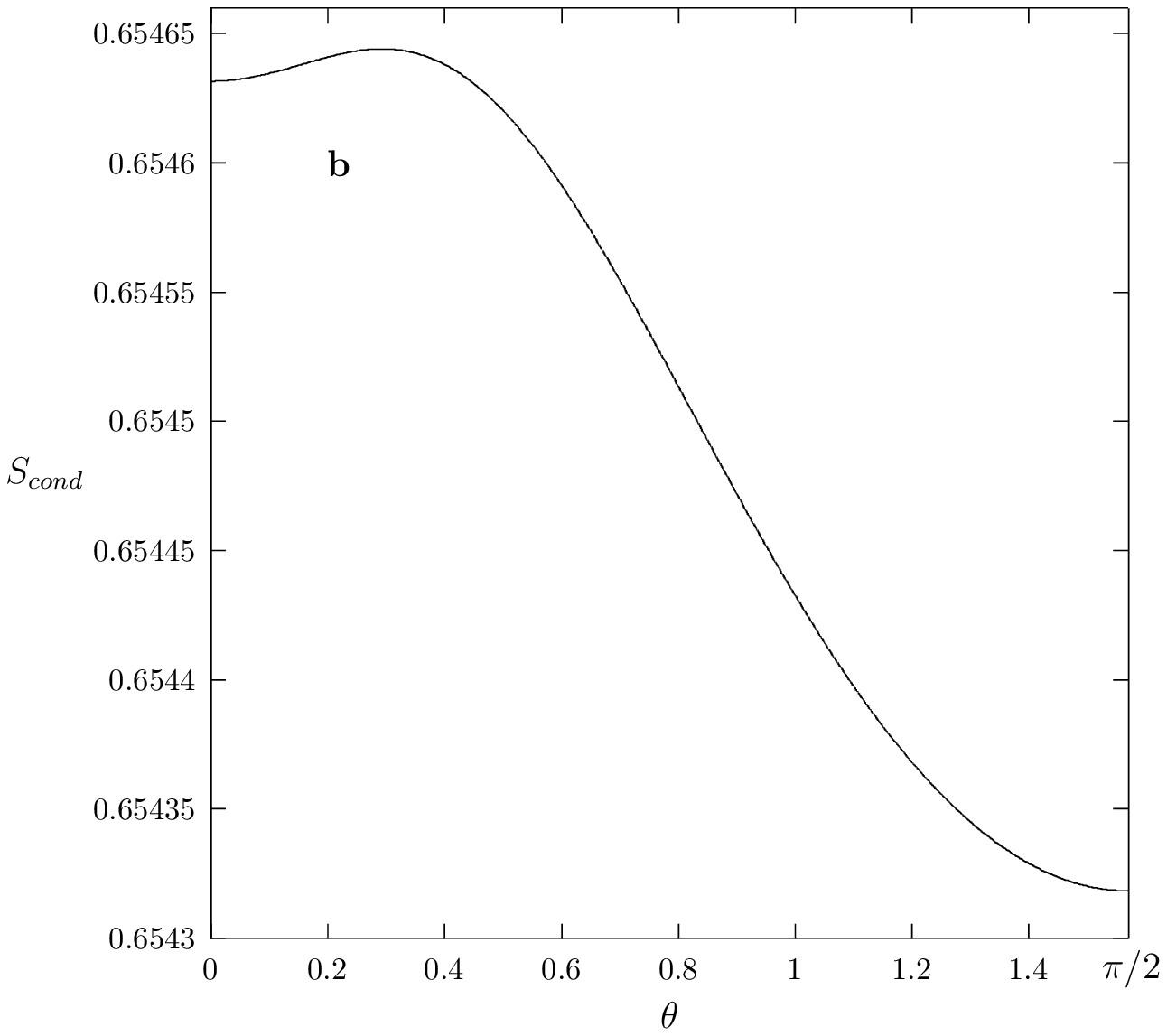,width=5.2cm}
\epsfig{file=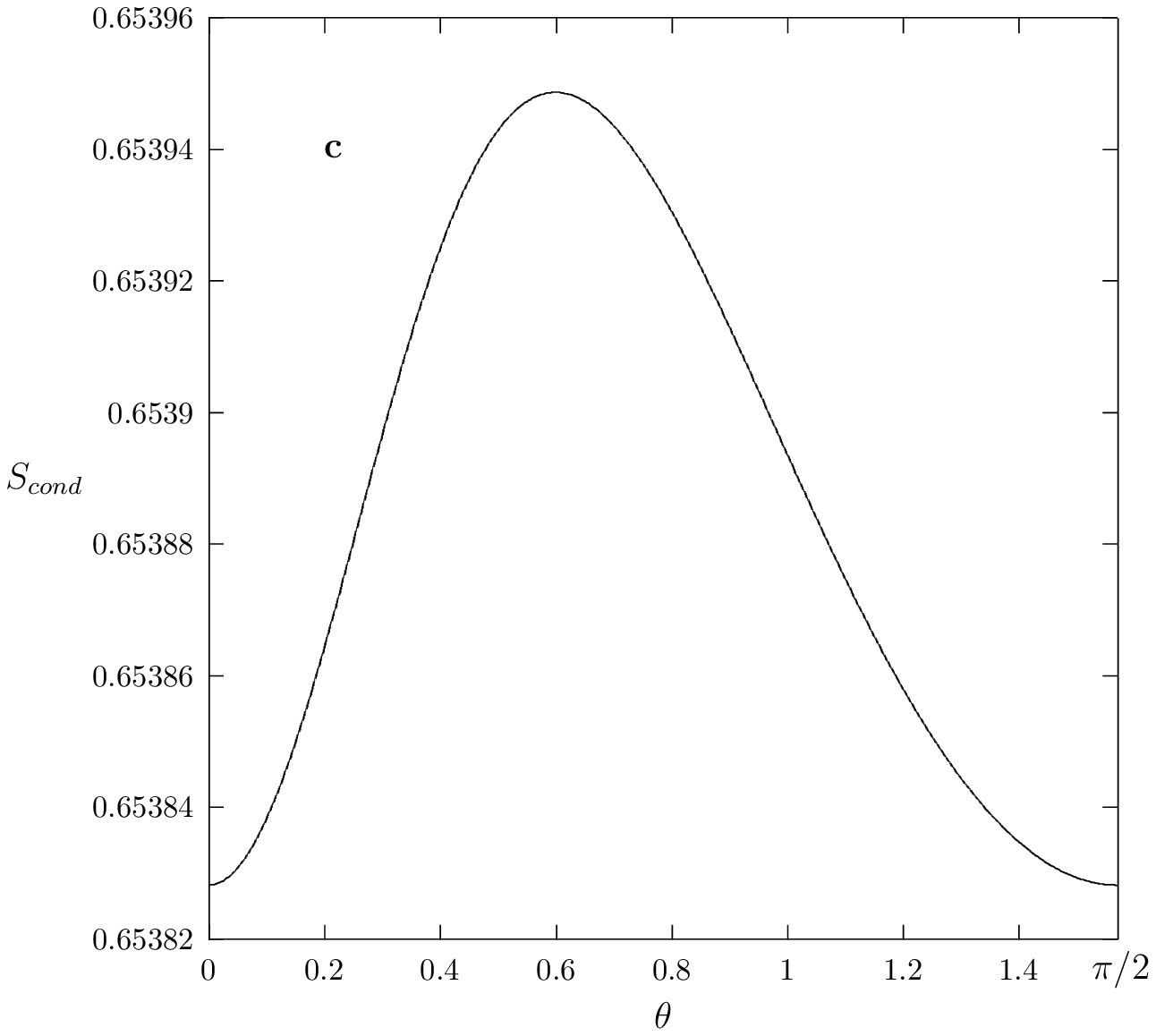,width=5.2cm}
\vspace{1cm}
\hspace{1cm}
\epsfig{file=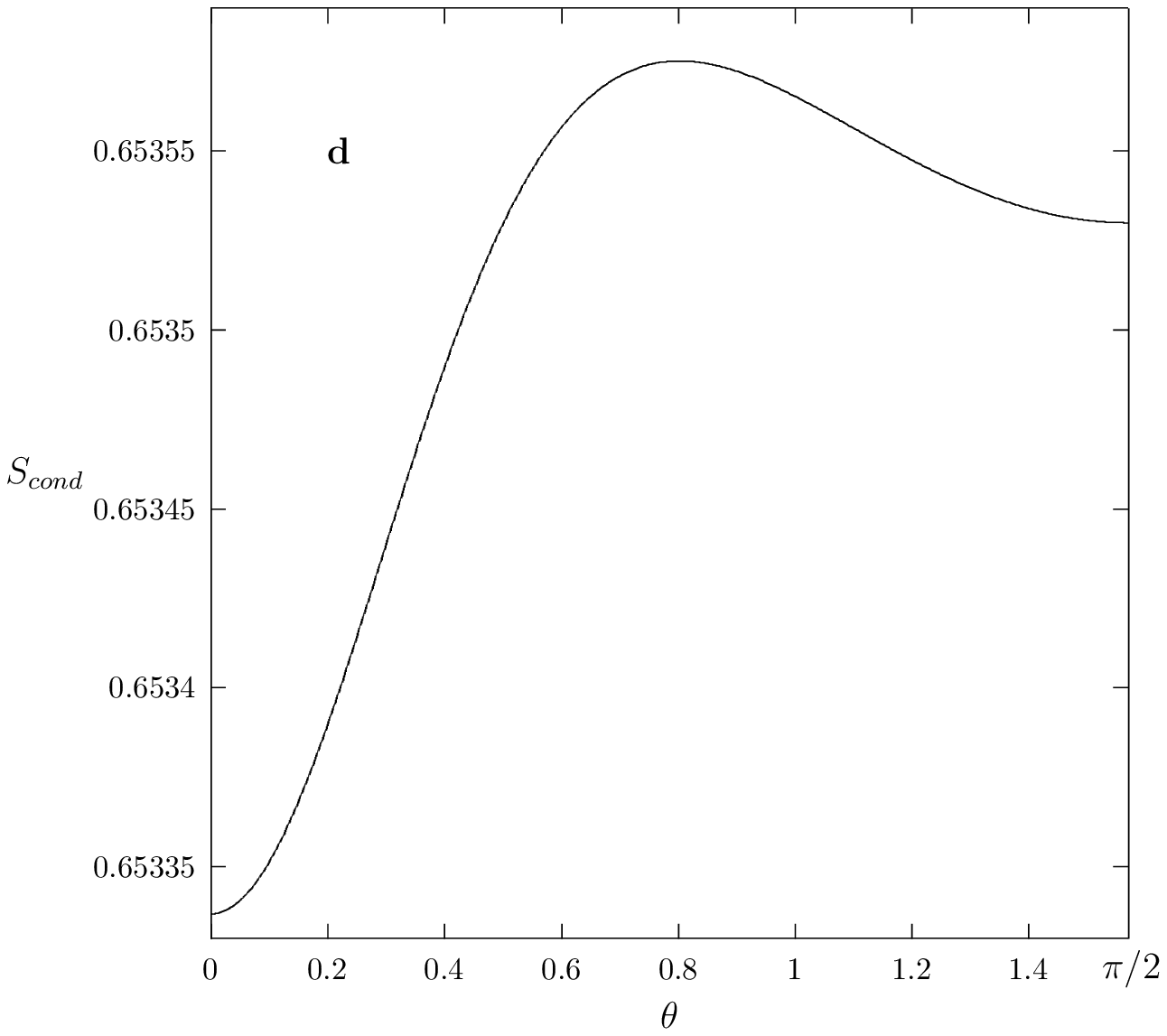,width=5.2cm}
\caption{
Conditional entropy $S_{cond}$ (in {\em bit}s) vs measurement angle $\theta$
by $c_3=0.34$, $c_1=0.14$, and $s_1=0.6477$~(a), 0.648~(b), $0.648\,435\,2971$~(c),
and 0.6487~(d).
Graphs show the evolution of intermediate maximum 
}
\label{fig:z034tr_b}
\end{center}
\end{figure}
The maximum as barrier separates two local minimuma at both bounds, $\theta=0$ and
$\pi/2$.

The maximum is born again thanks to the bifurcation mechanism but now the maximum at
$\theta=0$ splits into two maxima which are located upper and lower of this angle.
To find the bounds where the maximum exists, we again solve the transcendental
equations (\ref{eq:SII1}).
Numerical calculations give the band $0.647\,795\,5876\le s_1\le0.649\,123\,5832$.
First, that catches the eye, is that the 
width of this interval, $1.37\times10^{-3}$, is considerably larger (in an order)
than that for the minimum near the point $a$.
Fidelity between the end states of the interval is here equal to $F=99.998\,96\%$.
This quantity is smaller in comparison with the fidelity for the region with
minimum but it is again high too.

Largest excess of conditional entropy maximum is observed when
$S_{cond}(0)$ equals $S_{cond}(\pi/2)$ (see Fig.~\ref{fig:z034tr_b}(c)).
This is achieved at $s_1=0.648\,435\,2971$ (i.e., with high accuracy
at the middle of the interval with maximum).
Here $S_{cond}(0)=S_{cond}(\pi/2)=0.653\,828\,2081$~bit,
$S_{cond}^{max}=0.653\,948\,6436$~bit
(it occurs at $\theta=0.599\,427\,5584\approx34^{\circ}$)
and the relative excess of the maximum equals $0.018\%$ only. 
In the next section we find remarkably more suitable cases for observation of
intermediate conditional entropy maximum in an experiment.

Call attention that here, in contrast to the case with a minimum, the 0-boundary lies
in the $Q_{\pi/2}$ region whereas the $\pi/2$-boundary is in the $Q_0$ one.
So, the boundaries go over one another and this cannot be realized in practice.
Therefore, the only possibility for the discord is to change suddenly (abruptly)
the optimal measurement angle (i.e., to jump from $\pi/2$ to 0) at the crossing point
where $Q_{\pi/2}=Q_0$.
Above this point the discord is again returned to the old value
$Q_0=0.044\,231\,4345$~bit.

Evolution of quantum discord is shown in Fig.~\ref{fig:zq034tr_s}(1)
from $s_1=0$ to 0.67 along the strait line $c_1=0.14$ in the section $c_3=0.34$.
\begin{figure}[t]
\begin{center}
\epsfig{file=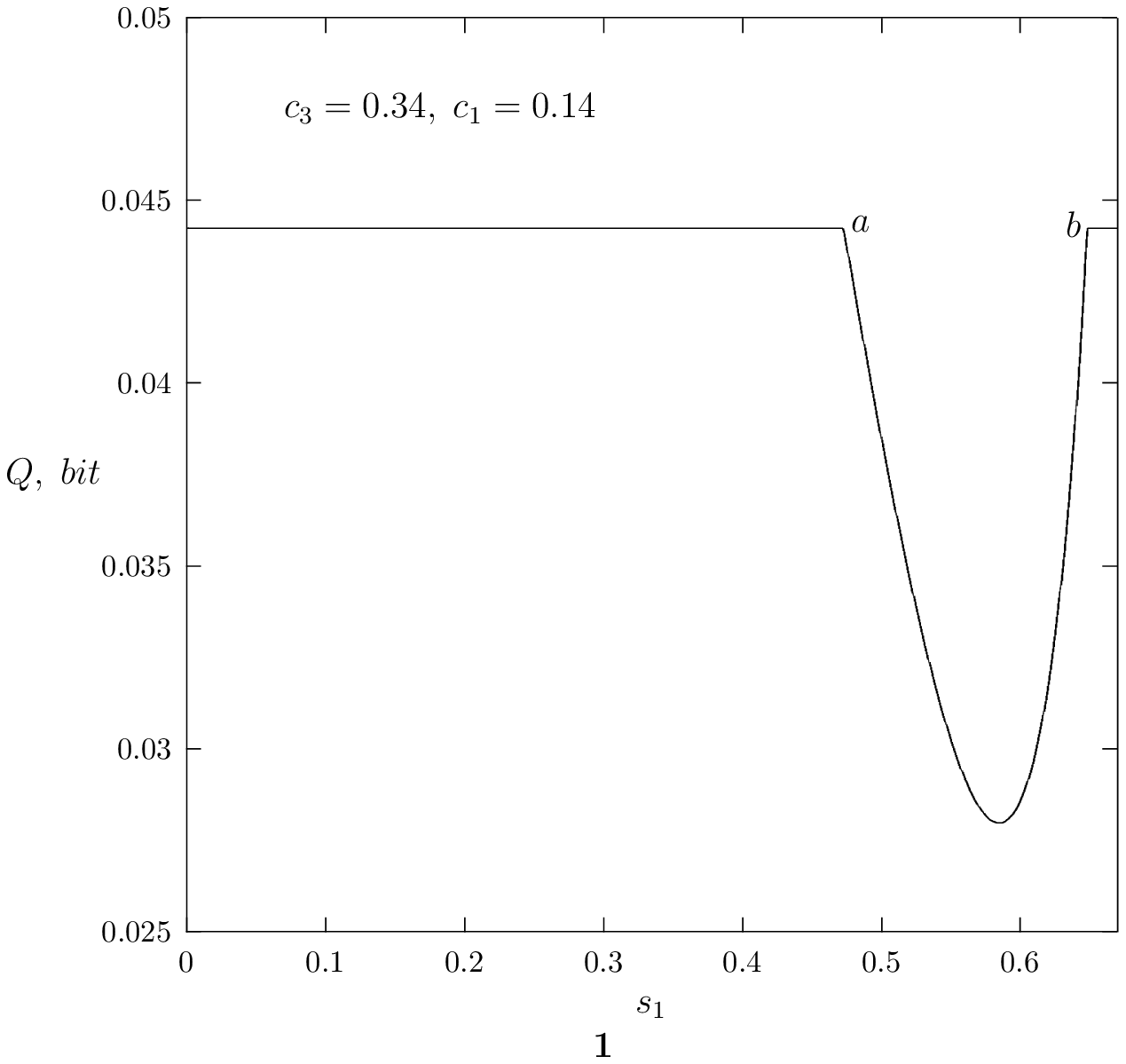,width=5.6cm}
\hspace{0.3cm}
\epsfig{file=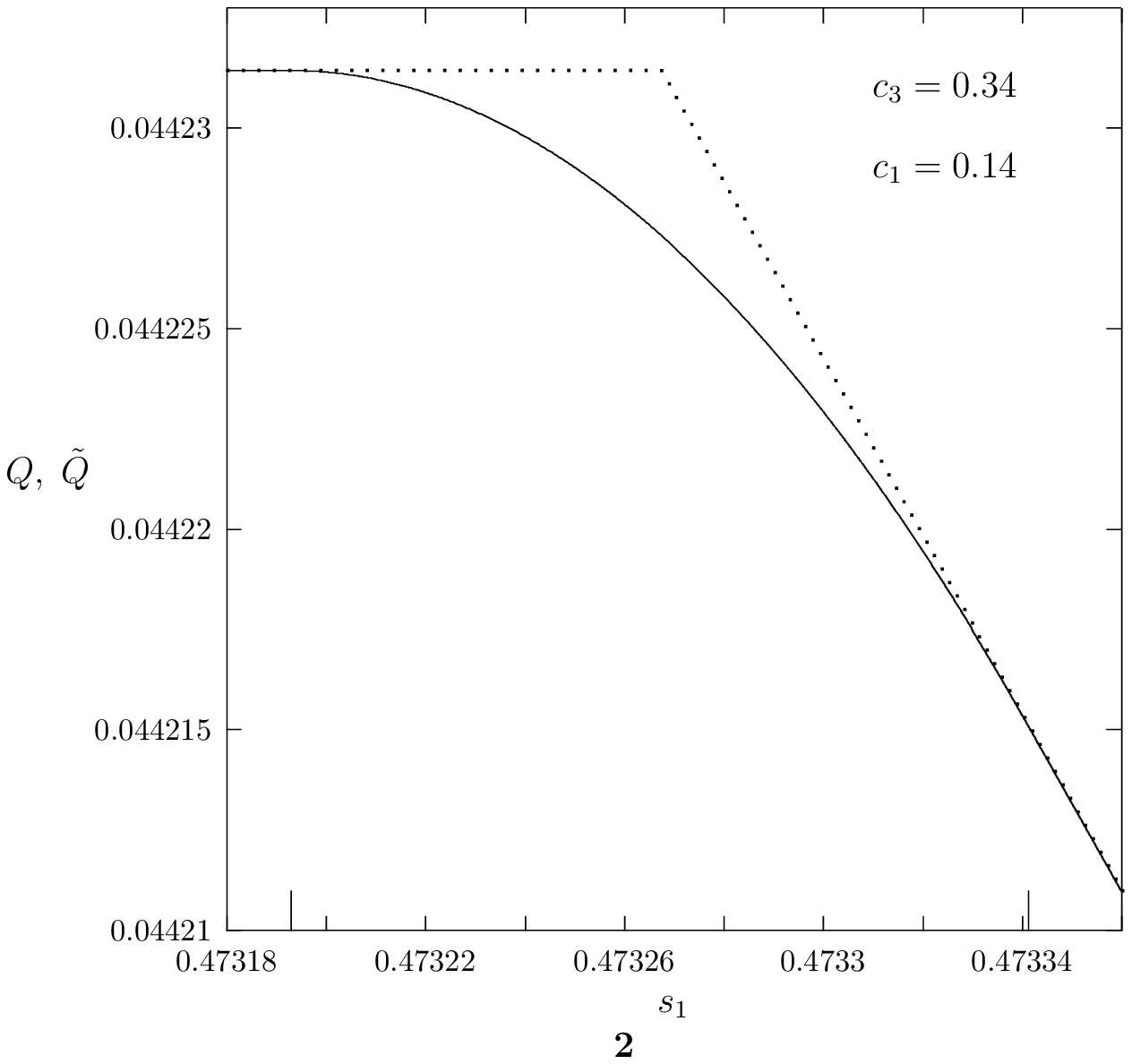,width=5.6cm}
\caption{
${\bf 1}$ shows the quantum discord $Q$ vs $s_1$ along the straight line
$c_1=0.14$ in the plane $c_3=0.34$.
${\bf 2}$ shows the true discord $Q$ (solid line) and false discord
${\tilde Q}$ (dotted line) near the point $a$.
Longer bars mark the bounds
$s_1^0=0.473\,192\,8814$ and $s_1^{\pi/2}=0.473\,341\,2570$
between which the $Q_{\theta^*}$ fracture exists
}
\label{fig:zq034tr_s}
\end{center}
\end{figure}
The curve has a minimum at the point with coordinates $(0.585, 0.027\,97)$.
According to Eqs.~(\ref{eq:JJzB}), the coupling constant $J_z$
changes its sing at $s_1=0.5997$ which correlates, more or less,
with the above location of minimum on the curve $Q$ upon $s_1$.

Figure~\ref{fig:zq034tr_s}(2) shows the behavior of quantum discord in the
vicinity of point $a$.
It is clearly seen how the $Q_{\theta^*}$ branch smoothly connects the branches
$Q_0$ and $Q_{\pi/2}$.
However, in spite of continuity and smoothness,
the quantum discord $Q$ experiences the sudden changes
at both points $s_1^0=0.473\,192\,8814$ and $s_1^{\pi/2}=0.473\,341\,2570$;
non-analytical behavior is displayed in higher derivatives.
This is in contrary to the statement of the authors~\cite{JYYL15} for
POVM-measurements.

We carefully inspected the behavior of conditional entropy in different domains
of tetrahedron ${\cal T}$.
Our conclusions are the following.
Regions with the interior maximum exist near all boundaries separating
the phases $Q_0$ and $Q_{\pi/2}$ for both $c_3<0$ and $c_3\ge0$.
These boundaries are shown in Fig.~\ref{fig:ph-d} by solid lines.
Regions with the interior minimum of conditional entropy are available only
by $c_3>0$ and exist as thin layers between the inner $Q_0$ region and $Q_{\pi/2}$
ones.
Their locations are schematically shown in Fig.~\ref{fig:ph-d} by double
solid-dotted lines and their structure is illustrated in Fig.~\ref{fig:z034dtr}.
One can say that here a ``sandwich''-like structure takes place.
Volume of the three-dimensional $Q_{\theta^*}$ layer gives an estimation of
corresponding states in the tetrahedron ${\cal T}$.
Numerical integration shows that the region $Q_{\theta^*}$ takes up the $0.08\%$
part of the volume of tetrahedron ${\cal T}$.

\section{Proposal for experiments}
\label{sect:Propos}
It is very tempting to discover a new phase in discordant materials or systems.
In the context being considered, this is the phase $Q_{\theta^*}$ where the conditional
entropy exhibits minimum inside the open interval $(0,\pi/2)$.
Despite the fact that the theory predicts existence of such regions
they don't seem to be experimentally resolved at present.
Instead, we suggest to detect another region which is very similar in the nature,
but where the conditional entropy has a local maximum in the interior of $(0,\pi/2)$.

A number of requirements seems to be needed, in our opinion,
for successful performing of the experiment.
First, minimal values of conditional entropy $S_{cond}$ must be more
that 0.5~bit.
Second, the excess of conditional entropy maximum should be not less than $1\%$
in comparison with the values of conditional entropy at the end points
$\theta=0$ and $\pi/2$.
And third (most difficult), the minimal sizes of region which contains the
conditional entropy maximum must be so large that the region could be
resolved on available experimental technique.

Trying to satisfy simultaneously the three named above criteria
(which, generally speaking, contradict each other) we are stopped here by
the following choice.
Let us take a cross section by the plane $c_3=-0.5$ that is drawn in 
Fig.~\ref{fig:ph-d}(a).
Figure~\ref{fig:z-05a} shows a part of the same phase diagram on which the
0- and $\pi/2$-boundaries are plotted.
\begin{figure}[t]
\begin{center}
\epsfig{file=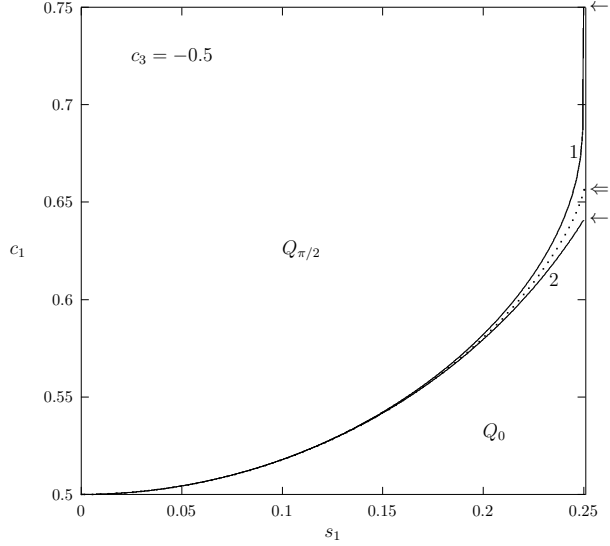,width=8cm}
\caption{
Fragment of a phase diagram in the section $c_3=-0.5$.
Dotted line is the boundary between $Q_0$ and $Q_{\pi/2}$ phases
whereas solid lines 1 and 2 are, respectively, the 0- and $\pi/2$-bounds
between which the region with conditional entropy maximum exists.
On the right vertical side $s_1=0.25$, two single arrows mark
the points $c_1^{\pi/2}=0.640\,668\,8666$ and $c_1^0=0.75$
and double arrow marks $c_1^\times=0.656\,390\,9127$ which are the end points of
$\pi/2$- and 0-bound and the ($Q_0-Q_{\pi/2}$)-boundary, respectively
}
\label{fig:z-05a}
\end{center}
\end{figure}
First of all notice that here the region between the 0- and $\pi/2$-boundaries is
clearly resolved visually.

The 0-boundary (curve 1 in Fig.~\ref{fig:z-05a}) goes across the $(s_1,c_1)$-plane
from the point ($0,0.5$) to the point $(0.25,c_1^0=0.75)$ while
the $\pi/2$-boundary (curve 2) runs from $(0,0.5)$ to its end point
$(0.25,c_1^{\pi/2}=0.640\,668\,8666)$.
Thus, the maximum distance in $c_1$-direction achieves the 'macroscopic'
value of $0.11$.
The fidelity of quantum states corresponding to the marked end points
on the right vertical side, $c_1^{\pi/2}$ and $c_1^0$ (see Fig.~\ref{fig:z-05a}),
equals $94.5\%$.
Such states might be easily distinguished in two-photon experiment where
the values $F=99.8(2)\%$ \cite{BSPBG13} and $F=99.8(1)\%$ \cite{Guo16} are achieved.
Unfortunately, as can be seen from Fig.~\ref{fig:z-05a}, the horizontal width
of the region under question rapidly decreases with increasing $c_1$
along the side $s_1=0.25$.

Consider the behavior of conditional entropy along the right side
of rectangle, i.e., when $c_3=-0.5$ and $s_1=0.25$ are held constant while $c_1$
varies from 0.5 to 0.75.
When $c_1$ is lower than the value of $\pi/2$-boundary the curve of $S_{cond}(\theta)$
has the only minimum at the angle $\theta=0$.
However, the situation is qualitatively changed for
$c_1\ge c_1^{\pi/2}=0.640\,668\,8666$.
Indeed, the maximum at $\theta=\pi/2$ decays in two ones,
the second minimum arises at the right end point, and
an inner maximum suddenly appears in the open interval $(0,\pi/2)$.
This process is shown on a set of graphs in Fig.~\ref{fig:zs-05tr}.
\begin{figure}[t]
\begin{center}
\epsfig{file=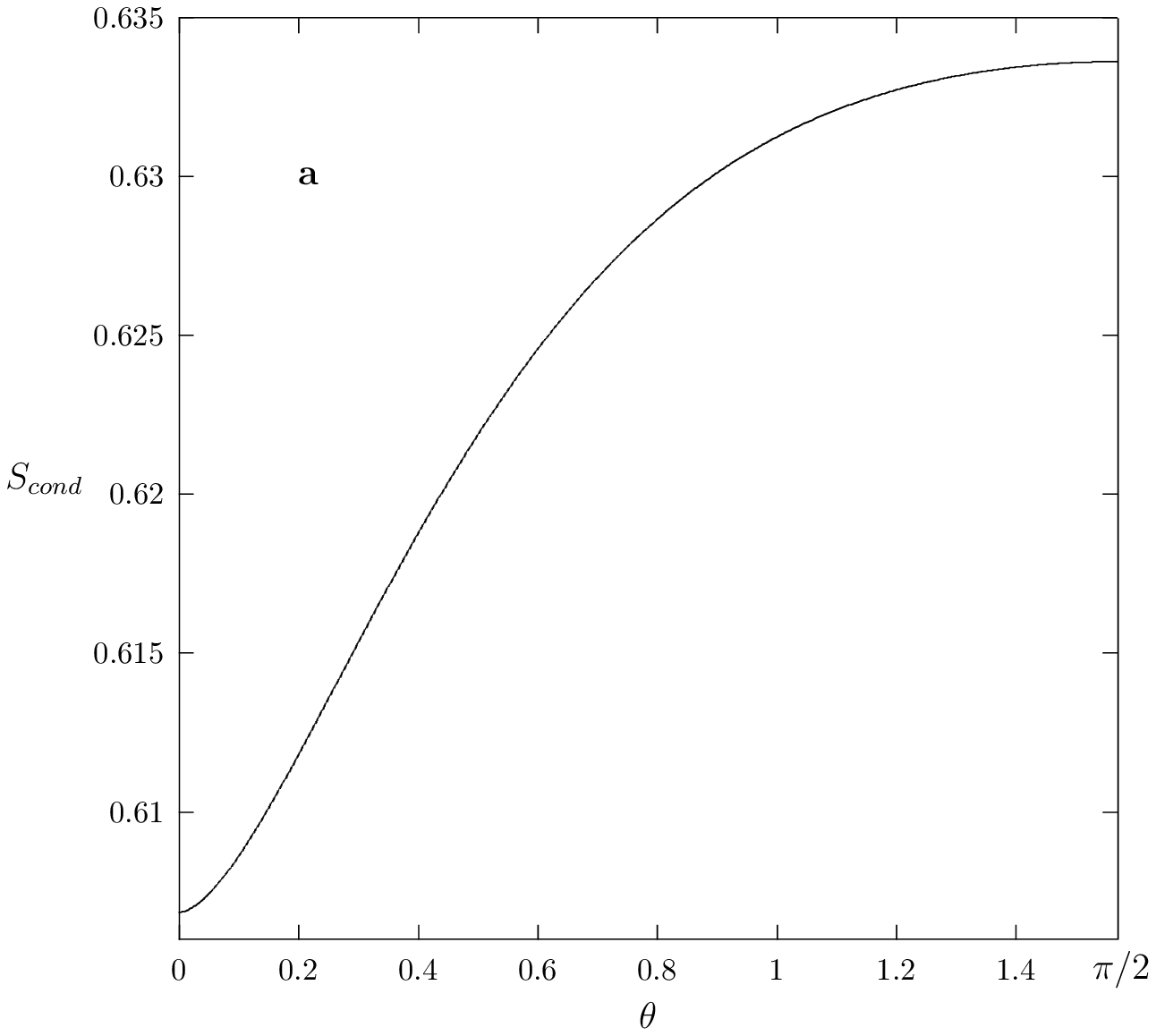,width=5.2cm}
\epsfig{file=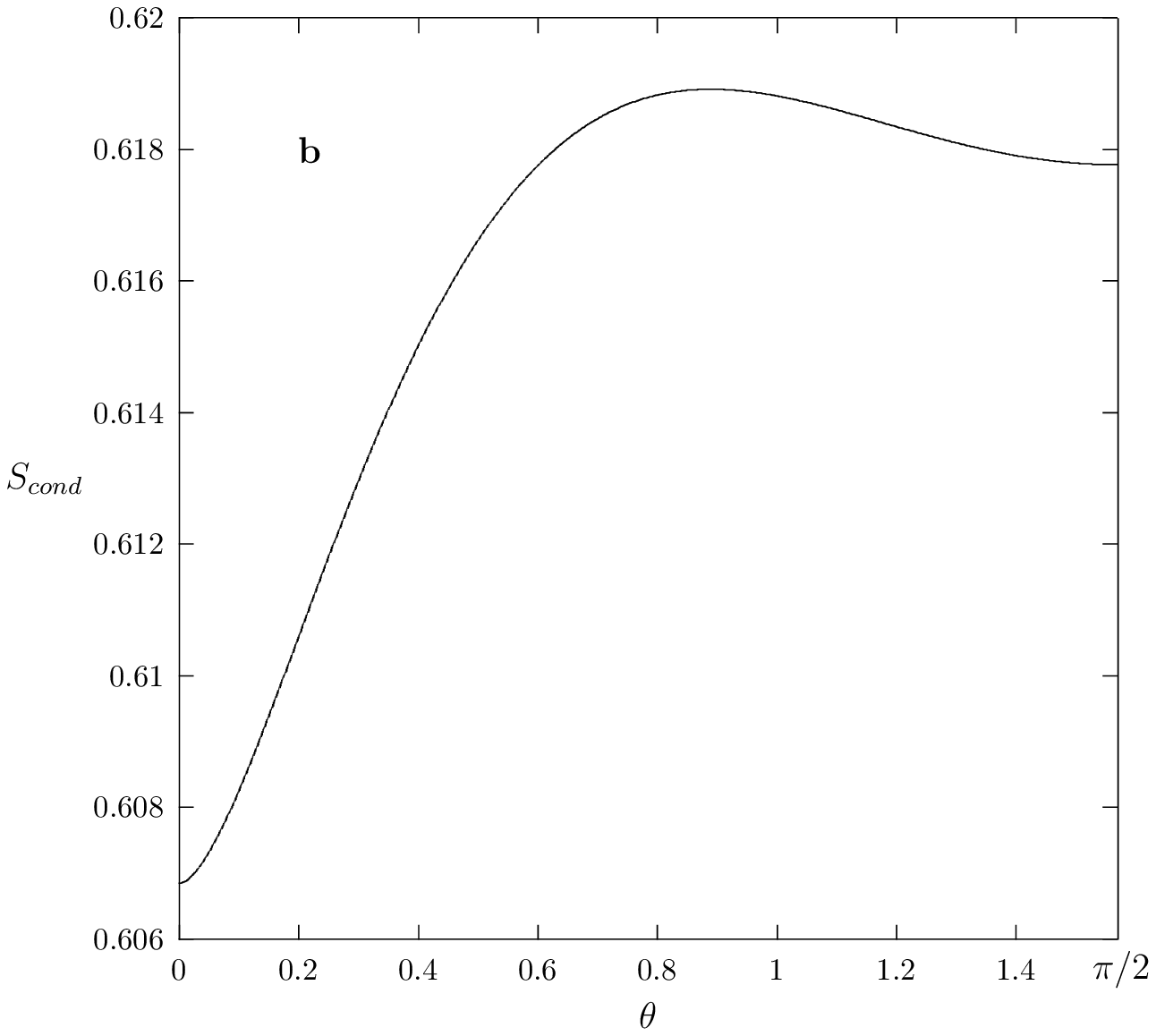,width=5.2cm}


\epsfig{file=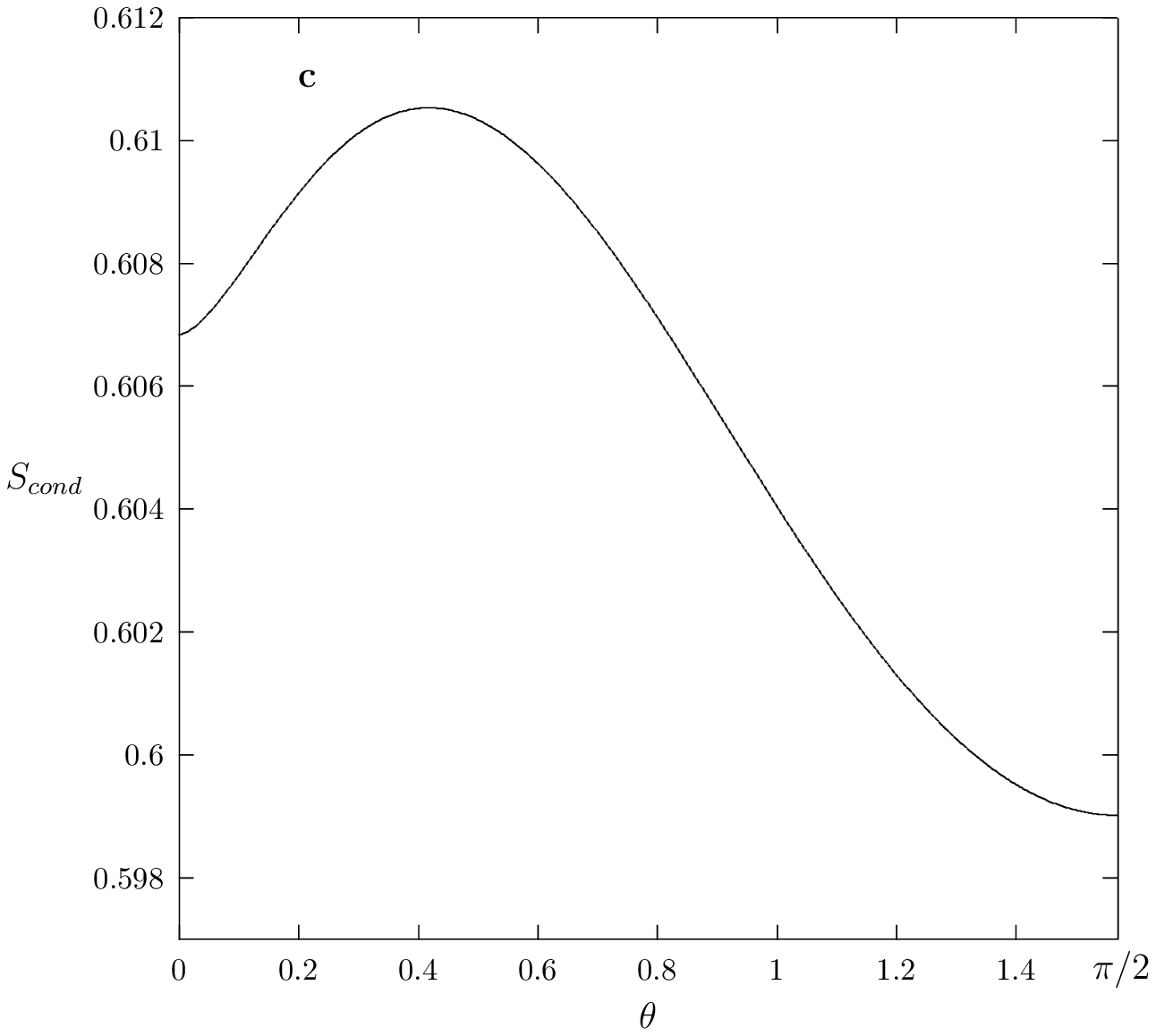,width=5.2cm}
\epsfig{file=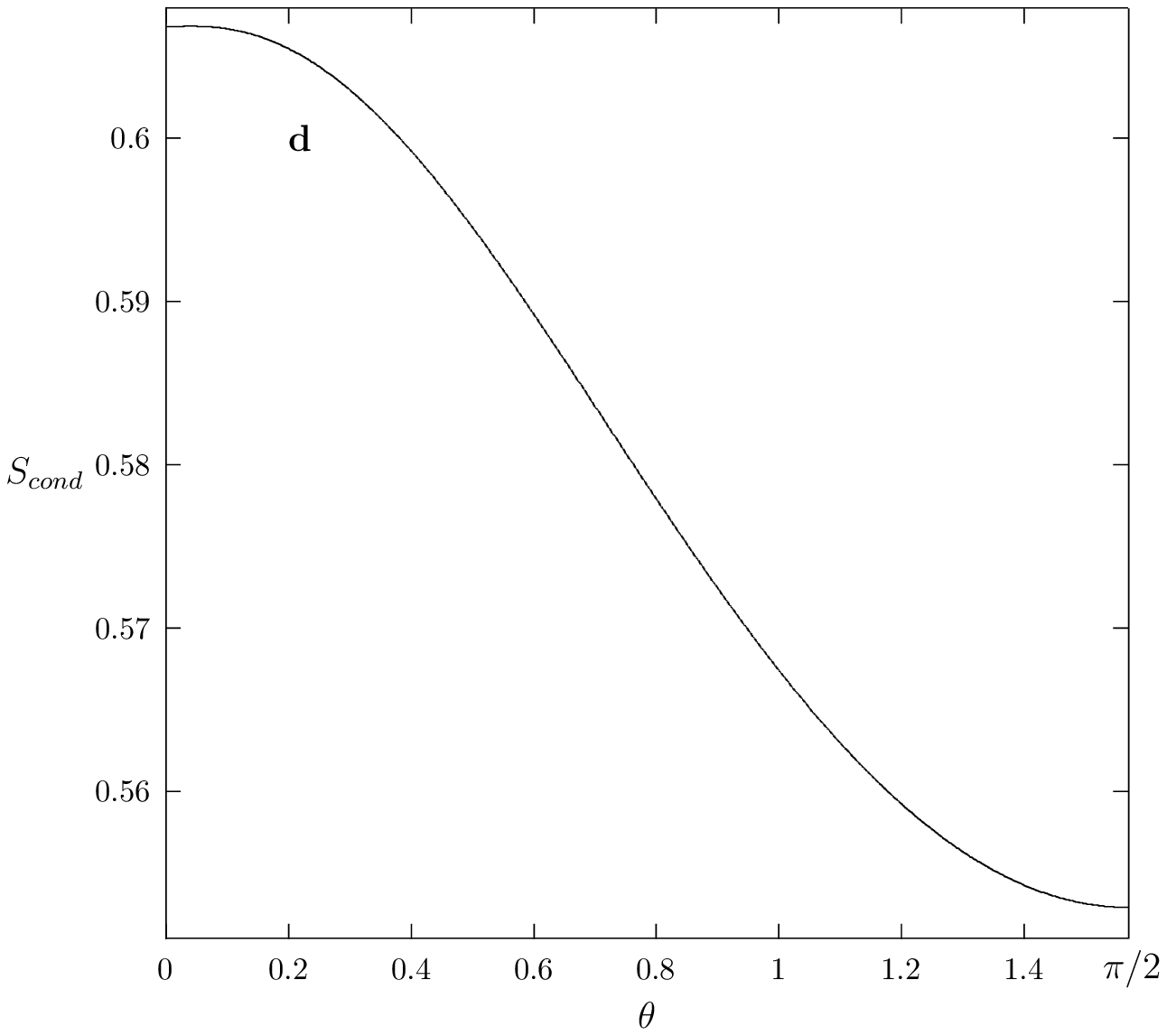,width=5.2cm}
\caption{
Graphs of average conditional entropy $S_{cond}(\theta)$, in bit units,
by $c_3=-0.5$, $s_1=0.25$, and
$c_1=0.633~ (a)$, $0.647~(b)$, $0.663~(c)$, and $0.7~(d)$.
Here one can see the appearance and disappearance of intermediate maximum
by moving along the vertical side of phase diagram in Fig.~\ref{fig:z-05a}
}
\label{fig:zs-05tr}
\end{center}
\end{figure}
The born maximum is kept dawn to $c_1=0.75$.

The fidelity of states on the side $s_1=0.25$ between the $\pi/2$-boundary,
$c_1^{\pi/2}=0.640\,668\,8666$, and the state, for example, at $c_1=0.7$
is $F=99.4\%$.
In principle, such states can be also resolved for photonics qubits.
Unfortunately, the excess of the maximum which first grows and achieves its maximal
value at the point $c_1^\times=0.656\,390\,9127$ then rapidly decreases.
So,  conditional entropy maximum by $c_1=0.7$ exists at the angle
$\theta=0.041\,813\,5775\approx2^\circ24^\prime$ but it is not seen
in Fig.~\ref{fig:zs-05tr}(d) because the value of maximum, $S_{cond}^{max}$,
exceeds the value of conditional entropy at $\theta=0$ on too small quantity:
$\delta S_{cond}^{max}=(S_{cond}^{max}-S_{cond}^0)/S_{cond}^{max}$
is $0.0049\%$ only.

As mentioned above, the largest excess of intermediate maximum occurs at the
phase transition point $c_1^\times$.
The corresponding curve of $S_{cond}(\theta)$ is presented in
Fig.~\ref{fig:zs-05tr_max}.
\begin{figure}[t]
\begin{center}
\epsfig{file=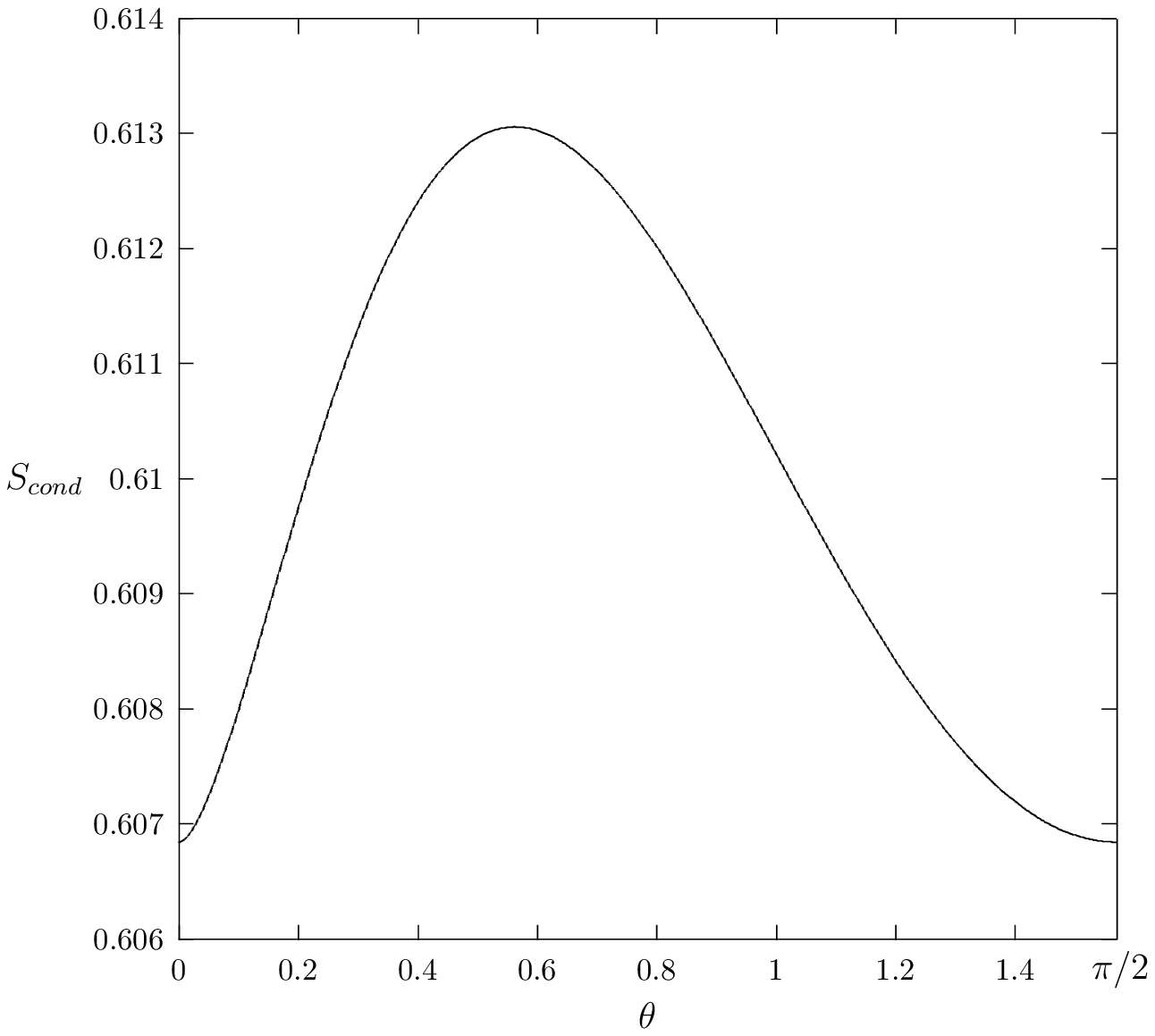,width=8cm}
\caption{
Average quantum conditional entropy
at the point $c_3=-0.5$, $s_1=0.25$, and $c_1=c_1^\times=0.656\,390\,9127$.
Maximum of conditional entropy is at the angle
$\theta_{max}=0.563\,770\,1781\simeq32^{\circ}$
and equals $S_{cond}^{max}=0.613\,058\,3056$~bit, i.e., relative excess
over the entropy values at $\theta=0$ or $\pi/2$ reaches 1.02\%
}
\label{fig:zs-05tr_max}
\end{center}
\end{figure}
The curve has typical bell-like shape.
It would be interesting to measure such a curve in a similar manner as it
has been done in the two-photon experiments \cite{XXLZZG10}\footnote
{
Due to the normalization of measurement angle $\theta$ taken
by the authors~\cite{XXLZZG10}, the curves of conditional entropy in their Fig.~2a
are essential only from zero to
45 degrees 
and therefore do not contain
interior local extrema what is in full agreement for the Bell-diagonal states.
}
and in \cite{XSLXGAFC13}.
In the latter paper, the scan precision of $\theta$ was $\pi/100$.

The maximum in Fig.~\ref{fig:zs-05tr_max} is wide enough, this being its
positive feature of course.
Values of $S_{cond}(\theta)$ at the end points $\theta=0$ and $\pi/2$ equal
$S_{cond}^{0,\pi/2}=0.606\,844\,1215$~bit.
Conditional entropy of two-qubit system can vary in the range from zero to one~bit
and therefore the given quantity is quite enough, in our opinion,
for measurements.
The excess of maximum is $1.02\%$ that also seems to be propitious circumstance
to perform the experiment.
The fidelity between the states at $c_1^{\pi/2}=0.640\,668\,8666$ and
$c_1^\times=0.656\,390\,9127$ equals $F=99.967\%$.

Similar results for different cross sections of tetrahedron ${\cal T}$
by planes $c_3=-0.8,\ldots,0.5$ for the side states with $s_1=(1+c_3)/2$
are collected in Table~\ref{tab:1}.
\begin{table}[t]
\caption{$S_{cond}$ at $\theta=0$ or $\pi/2$,
relative excess of maximum, and fidelity between $c_1^\times$ and
$c_1^{\pi/2}$ surface states (on the face $v_1v_3v_4$, see Fig.~\ref{fig:z_xxz-m2})
vs different values of cross-section parameter $c_3$; here $s_1=(1+c_3)/2$
\upshape\upshape}
\label{tab:1}
\begin{tabular}{clll}
\hline\noalign{\smallskip}
$c_3$ & $S_{cond}^{0,\pi/2}$, bit & Excess & Fidelity \\
\noalign{\smallskip}\hline\noalign{\smallskip}
$-0.8$ & $0.376\,221\,1396$ & $1.1\%$ & $0.999\,89$ \\
$-0.5$ & $0.606\,844\,1215$ & $1.02\%$ & $0.999\,67$ \\
$-0.1$ & $0.694\,226\,2757$ & $0.77\%$ & $0.999\,45$ \\
$0$ & $0.668\,721\,8755$ & $0.71\%$ & $0.999\,42$ \\
$0.1$ & $0.673\,581\,6250$ & $0.64\%$ & $0.999\,41$ \\
$1/3$ & $0.601\,606\,7457$ & $0.471\%$ & $0.999\,43$ \\
$0.5$ & $0.517\,713\,6813$ & $0.35\%$ & $0.999\,49$ \\
\noalign{\smallskip}\hline
\end{tabular}
\end{table}
As seen from the table, the relative excess of conditional entropy maximum
decreases monotonically with increasing $c_3$ whereas $S_{cond}^{0,\pi/2}$ reaches
its largest value for $c_3=-0.1$ while the fidelity is minimal when $c_3=0.1$.
Notice the values of fidelity can be somewhat reduced if one takes the states in the
$c_3$-direction.

\section{Conclusions}
\label{sect:Concl}
In the present paper, the discord phase diagram of X states with $s_1=s_2$ and
$c_1=c_2$ has been built up.
Location and exact boundaries for the discordant $\theta$-fraction have been
established.

We have also found different types of conditional entropy behavior
varying the arguments $s_1$, $c_1$, and $c_3$ belonging to their full
domain of definition ${\cal T}$.
Firstly, the curve of function $S_{cond}(\theta)$ can be a straight line from zero
up to $\pi/2$ (type $I$).
This situation is realized, for example, by $s_1=c_1=c_3=0$ or, say, by $s_1=0$
and $c_1=\pm c_3$.
As follows from Eq.~(\ref{eq:Sconds1c1c3}), here
$S_{cond}(\theta;0,c_1,\pm c_1)=\ln2$~nat (=1~bit) for $\forall\,\theta\in[0,\pi/2]$.
Secondly, there are monotonically increasing or, v.v, decreasing behaviors
in the whole closed interval $[0,\pi/2]$ (types $II$ and $III$, respectively).
Such shapes of conditional entropy lead to the $Q_0$ and $Q_{\pi/2}$ fractions,
respectively.
In both cases, there is only one inflection point inside the interval $(0,\pi/2)$.
Thirdly, the conditional entropy can have a local minimum inside the interval
$(0,\pi/2)$.
This is the $IV$ type of its behavior.
In such a case, the intermediate phase of quantum discord $Q_{\theta^*}$ arises.
And finally, we have found regions in the space of X states where the conditional
entropy exhibits a local maximum in the interior of interval $(0,\pi/2)$.
This is the $V$ type of conditional entropy behavior.

Both the minimum and maximum suddenly appear and disappear at the end points of
interval $[0,\pi/2]$ through a bifurcation mechanism.
Boundaries for such regions satisfy the same equations (\ref{eq:SII1}).
To distinguish the types of regimes ($IV$ or $V$ according to the above presented
classification) it is needed to perform an additional analysis, namely it is required
to study the shapes of conditional entropy between the found boundaries.

The region with conditional entropy minimum (the $\theta$-phase of discord) is very tiny
and possibilities are not seen to observe it experimentally.
On the other hand, the region with conditional entropy maximum is remarkably larger
and there exists hope to obtain the experimental evidence for its presence.

The branches $Q_0$, $Q_{\pi/2}$, and $Q_{\theta^*}$ are
analytic dependencies but the quantum discord at boundaries between them
experiences sudden changes which can be observed in their higher derivatives.

As a whole one should say the following.
Evaluation of quantum discord for X states with nonzero local Bloch vectors is far
from completeness and perfection of Luo's formula.
Above all, the unimodality of conditional entropy stays an open question.
Discussion of this problem is given in the Appendix.
Further, we do not know where the regions with interior minimum and maximum are located
in the full five-parameter X space and what their characteristics
(values of extrema, sizes of such regions, etc.) are.
One can hope that the answers to these and other questions will be found in
the future investigations.


\vspace{-10mm}
\section*{}
{\bf Acknowledgment}\ The research was supported by the RFBR Grant 15-07-07928.

\section*{Appendix. Unimodality hypothesis for X states}
\label{sect:App}
Here, we formulate the unimodality hypothesis for the average
entropy of post-measurement states (conditional entropy)
and entropy of weighted average post-measurement state (entropy after measurement)
of two-qubit X states.
These entropies enter the expressions of quantum discord and
one-way quantum deficit, respectively.

Start with definitions and enumerate some known facts in this mathematical field.

\medskip
{\em Definition of strong unimodality}.
A function $f(x)$ is a unimodal function in the interval $[a,b]$ if for some value 
$x_m\in[a,b]$, it is monotonically increasing for $x\le x_m$ and monotonically
decreasing for $x\ge x_m$.
In that case, the maximum value of $f(x)$ is $f(x_m)$ and there are no other
local maxima.

\medskip
{\em Definition of weak unimodality}.
A function $f(x)$ is a weakly unimodal function in the interval $[a,b]$ if there exists
a value $x_m\in[a,b]$ for which it is weakly monotonically increasing for $x\le x_m$
and weakly monotonically decreasing for $x\ge x_m$.
In that case, the maximum value $f(x_m)$ can be reached for a continuous range
of values of $x$.
\medskip

Analogous definitions are given for the minimum.
Simplest examples provide the polynomial function of second degree,
$f(x)=ax^2+bx+c$.

Proving unimodality for nontrivial functions is often hard.
A general method based on derivatives is discussed in Ref.~\cite{B13}.

Obviously, the function is a unimodal one if it is convex/concave in the interval
$[a,b]$.
This is a sufficient condition.
Further, $f$ is unimodal if there is one to one differentiable mapping $x=g(y)$
such that $f(g(y))$ is convex/concave.
The latter allows to prove the unimodality property in some cases.

We consider two functions of one variable $x\in[0,1]$ with five real parameters
$p_1,\ldots,p_5$,
\begin{eqnarray}
   \label{eq:fxp1-5a1}
   &&f_1(x;p_1,p_2,p_3,p_4,p_5)=-h_2\biggl(\frac{1+p_2x}{2},\frac{1-p_2x}{2}\biggr)
	 \nonumber\\
	 &&+h_4\biggl(\frac{1+p_2x+\sqrt{r_1}}{4},\frac{1+p_2x-\sqrt{r_1}}{4},\frac{1
	 -p_2x+\sqrt{r_2}}{4},\frac{1-p_2x-\sqrt{r_2}}{4}\biggr)\ ({\rm A}1) 
	 \nonumber
\end{eqnarray}
and
\begin{eqnarray}
   \label{eq:fxp1-5a2}
   &&f_2(x;p_1,p_2,p_3,p_4,p_5)=
	 \nonumber\\
	 &&+h_4\biggl(\frac{1+p_2x+\sqrt{r_1}}{4},\frac{1+p_2x-\sqrt{r_1}}{4},\frac{1
	 -p_2x+\sqrt{r_2}}{4},\frac{1-p_2x-\sqrt{r_2}}{4}\biggr)\ ({\rm A}2) 
	 \nonumber
\end{eqnarray}
with $r_{1,2}=(p_1\pm p_5x)^2+4w^2(1-x^2)$ and $w=(|p_3+p_4|+|p_3-p_4|)/4$.
\begin{figure}[t]
\begin{center}
\epsfig{file=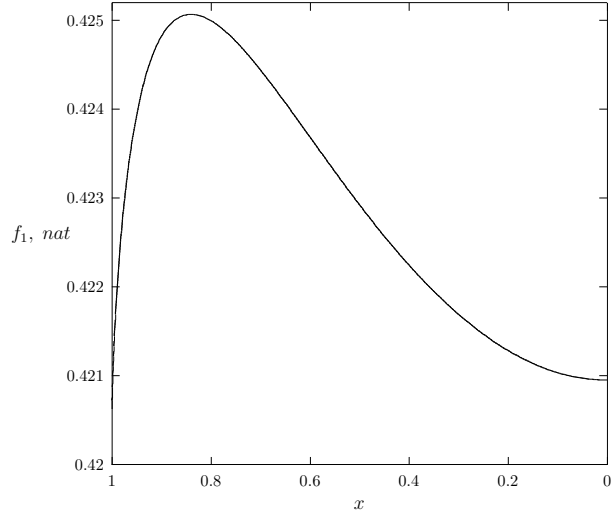,width=8cm}
\caption{
Behavior of function $f_1$, Eq.~(A1), vs $x$ by $p_1=p_2=0.25$, $p_3=p_4=0.656$,
and $p_5=-0.5$
(cf. with Fig.~\ref{fig:zs-05tr_max})
}
\label{fig:zz2}
\end{center}
\end{figure}
These functions, $f_1(x)$ and $f_2(x)$, correspond to the conditional entropy
\cite{Y15} and entropy after measurement \cite{YWF16,YF16}\footnote{
Results of papers \cite{YWF16,YF16} are restricted to $p_3+p_4\ge0$ and $p_3-p_4\ge0$
(one needs every time to reduce the initial X density matrix to a form with real
nonnegative off-diagonal entries)
whereas our expressions (A1) and (A2) are automatically valid in the whole seven-parameter
domain of X states thanks to the quantity $w$ \cite{Y14,Y14a,Y15}.
}, respectively.

In the expressions~(A1) and (A2),
$h_2(x_1,x_2)=-x_1\log{x_1}-x_2\log{x_2}$
with additional condition $x_1+x_2=1$
and 
$h_4(x_1,x_2,x_3,x_4)=-x_1\log{x_1}-x_2\log{x_2}-x_3\log{x_3}-x_4\log{x_4}$
with condition $x_1+x_2+x_3+x_4=1$
are, respectively, the binary and quaternary entropy Shannon functions;
$0\le h_2\le1$~bit and $0\le h_4\le2$~bits.
These functions are convex \cite{CT91,K09}.
It is also known that the sum of convex functions is convex again but
the difference is not in general.

\medskip
{\em Conjecture}. The functions $f_1(x)$ and $f_2(x)$ for every choice of parameters
$p_1,\ldots,p_5$ for which all arguments of Shannon functions are non-negative
can have at most only one local extremum (minimum or maximum) in the open interval
$x\in(0,1)$.
\medskip

It is required to prove or refute this proposition.
Numerical calculations \cite{Y14,Y14a,Y15,YWF16} give evidence which supports the
above conjecture.
Figures~\ref{fig:zz2} and \ref{fig:zq0615a} show the dependencies of $f_1(x)$ and
$f_2(x)$ for specific choice of parameters $p_1,\ldots,p_5$.
Notice that the used mapping $x=\cos{\theta}$ has led, as seen from the figures,
to convexity/concavity of both functions for $x\ge x_m$ unlike the same
quantities
upon the angle $\theta$.
So, it is enough to find an additional mapping which provides
convexity/concavity below
the extremal points $x_m$.
\begin{figure}[t]
\begin{center}
\epsfig{file=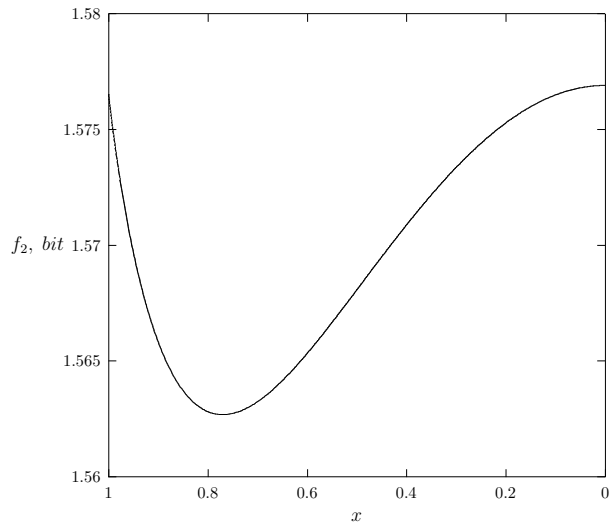,width=8cm}
\caption{
Behavior of function $f_2$, Eq.~(A2), vs $x$ by $p_1=p_2=0.385$, $p_3=p_4=-0.615$,
and $p_5=-0.23$
}
\label{fig:zq0615a}
\end{center}
\end{figure}



\end{document}